\newif\ifsubmode
\submodefalse

\newif\ifprintfig
\printfigtrue

\newif\ifemulate
\emulatetrue
\ifsubmode
\documentclass[12pt,preprint]{aastex}
\received{}
\accepted{}
\journalid{}
\articleid{}
\else
   \documentclass{emulateapj}
   \submitted{\it Accepted for publication in ApJ}
\fi
\usepackage{multirow,color,natbib,wrapfig,ulem}
\usepackage{placeins}
\def\spose#1{\hbox to 0pt{#1\hss}}
\def\simlt{\mathrel{\spose{\lower 3pt\hbox{$\mathchar"218$}}
     \raise 2.0pt\hbox{$\mathchar"13C$}}}
\def\simgt{\mathrel{\spose{\lower 3pt\hbox{$\mathchar"218$}}
     \raise 2.0pt\hbox{$\mathchar"13E$}}}

\shorttitle{Reduced Masses of Satellites}
\shortauthors{Zolotov et al.}

\begin{document} 

\title{Baryons Matter: Why Luminous Satellite Galaxies Have Reduced Central Masses}

\author{Adi\ Zolotov\altaffilmark{1}, Alyson\ M.\ Brooks\altaffilmark{2}, Beth Willman\altaffilmark{3}, Fabio Governato\altaffilmark{4}, Andrew Pontzen\altaffilmark{5}, Charlotte Christensen\altaffilmark{6}, Avishai Dekel\altaffilmark{1}, Tom Quinn\altaffilmark{4}, Sijing Shen\altaffilmark{7}, James Wadsley\altaffilmark{8}}

\altaffiltext{1}{Racah Institute of Physics, The Hebrew University,
  Jerusalem, Israel 91904; adizolotov@gmail.com}
\altaffiltext{2}{Department of Astronomy, University of Wisconsin-Madison, 475 N. Charter St., Madison, WI 53706 USA}
\altaffiltext{3}{Department of Astronomy, Haverford College, 370 Lancaster Ave, Haverford, PA 19041 USA}
\altaffiltext{4}{Astronomy Department, University of Washington, Box 351580, Seattle, WA 98195 USA}
\altaffiltext{5}{Oxford Astrophysics, University of Oxford, Denys Wilkinson Building, Keble Road, Oxford OX1 3RH}
\altaffiltext{6}{Department of Astronomy/Steward Observatory, 933 North Cherry Ave, Tucson, AZ 85721 USA}
\altaffiltext{7}{Department of Astronomy and Astrophysics, University of California, Santa Cruz, CA 95064, USA}
\altaffiltext{8}{Department of Physics and Astronomy, McMaster University, Hamilton, Ontario L88 4M1, Canada}
\date{\today}
\begin{abstract}

Using high resolution cosmological hydrodynamical simulations of Milky
Way-massed disk galaxies, we demonstrate that 
supernovae feedback and tidal stripping lower the central masses of bright
($-15 < M_V < -8$) satellite galaxies. These simulations resolve high
density regions, comparable to giant molecular clouds, where stars
form.  This resolution allows us to adopt a prescription for H$_2$
formation and destruction that ties star formation to the presence of
shielded, molecular gas.  
Before infall, supernova feedback from the clumpy, bursty star formation captured by
this physically motivated model leads to reduced dark matter (DM)
densities and shallower inner density profiles in the massive
satellite progenitors (M$_{vir} \ge 10^9 M_{\odot}$, M$_{*} \ge 10^7
M_{\odot}$) compared to DM-only simulations.  The progenitors of the lower 
mass satellites are unable to maintain bursty star formation histories, due 
to both heating at reionization and gas loss from initial star forming
events, preserving the steep inner density profile predicted by DM-only simulations.
After infall, gas stripping from satellites reduces the total
central masses of SPH satellites relative to DM-only satellites.
Additionally, enhanced tidal stripping after infall due to the
baryonic disk acts to further reduce the central DM densities of the
luminous satellites.  Satellites that enter with cored DM halos are 
particularly vulnerable to the tidal effects of the disk, 
exacerbating the discrepancy in the central
masses predicted by baryon+DM and DM-only simulations. 
We show that DM-only
simulations, which neglect the highly non-adiabatic evolution of baryons 
described in this work, produce denser satellites with larger central velocities.  We
provide a simple correction to the central DM mass predicted for
satellites by DM-only simulations. We conclude that DM-only
simulations should be used with great caution when interpreting
kinematic observations of the Milky Way's dwarf satellites.

\end{abstract}

          
\section{INTRODUCTION}\label{intro_sec}

The favored Cold Dark Matter (CDM) cosmological model has been
successful in reproducing many large scale observable properties of
the Universe \citep[e.g.,][]{Efstathiou1992, Riess1998, Spergel2007}.
The CDM model, however, still faces many challenges from observations
of galaxies on small scales.  The most well known of these problems
has been termed the ``missing satellite problem'',
since CDM-based models predict orders of 
magnitude more dark matter subhalos within the virial radii of Milky Way-massed galaxies 
than are observed as luminous satellites of such systems \citep{Moore1999, klypin1999, Wadepuhl2011}. 
Another aspect of the missing satellites problem is the discrepancy
between the masses of the most massive predicted subhalos and the
most massive observed satellites \citep{Moore1999, klypin1999,
 Boylan-kolchin2011, Boylan-kolchin2012}. 
Simulated $\sim 10^{12} M_{\odot}$ halos consistently have
several subhalos that are too massive and too dense to host the most
luminous dwarf spheroidal (dSph) satellites of the Milky Way
\citep{Boylan-kolchin2011, Boylan-kolchin2012, Wolf2012, Hayashi2012}.  
This appears to provide a new challenge for the CDM paradigm on small scales,
because the most massive subhalos of an L$^{\star}$ galaxy should not be
devoid of stars.

The tension between the predicted and observed inner densities of the 
Milky Way's dSph satellites is reminiscent of the longstanding tension
between the predicted and observed shapes of the central density profiles 
of galaxies (known as the ``core/cusp problem'').  The steep inner density
profiles and concentrations predicted for dark matter halos and their
satellites \citep{Navarro1997a, Lia2000, Dekel2003a, Dekel2003b, Reed2005, 
Springel2008, Madau2008, Navarro2010, Maccio2009} are inconsistent with 
those observed in isolated field galaxies \citep{Persic1996, vandenbosch2000, 
deblok2001, deblok2002, Simon2003, Swaters2003, Weldrake2003, Kuzio2006, 
Salucci2007, Gentile2007, Spano2008, Trachternach2008, deblok2008, 
Donato2009, Oh2011, Delpopolo2012}, and observed in satellites 
\citep{Kleyna2002, Kleyna2003, Mashchenko2005, Goerdt2006, Strigari2006, 
Gilmore2007, Walker2009, Strigari2010, Walker2011, Jardel2012, Wolf2012, 
Hayashi2012, Salucci2012}.  This inconsistency is independent of whether 
the density slope in simulations follows log($\rho$) $\propto$ $\gamma$ log(radius), 
with 1.0 $< \gamma <$ 1.5 \citep[i.e., a NFW profile,][]{Navarro1997a}, or 
a power-law slope \citep[i.e., an Einasto profile,][]{Navarro2010}.
 
Baryonic processes have often been proposed to address the apparent 
discrepancies between observations and the predictions of DM-only simulations. 
Within the CDM paradigm, the missing satellite problem is likely
reconciled through reionization, which is expected to suppress star formation 
in subhalos with maximum circular velocity, $v_c$, at infall $<$ 30 km/s   
\citep{Quinn1996, Thoul1996, Navarro1997, Gnedin2000, Hoeft2006, Okamoto2008, 
Madau2008}, and supernova feedback further suppressing star formation at 
the more massive subhalo end \citep{Dekel1986, Benson2002, Dekel2003c,
Governato2007, Busha2010}.  When combined with observational  
incompleteness effects \citep{Simon2007,Tollerud2008, Walsh2009, Koposov2009}, 
it is likely that baryonic effects can bring the predicted number of subhalos 
in line with the observed number of satellites.  Yet even if the correct number of satellites 
can be reproduced, CDM still predicts that the most massive satellites today  
are more massive and more dense than observed for Milky Way satellites 
\citep{Boylan-kolchin2012, Wolf2012, Hayashi2012}. 

Baryons have also been invoked to reconcile the cusp/core problem in CDM.
The predicted cuspy profiles of dark matter halos may not be physically 
realistic because the effect of energetic feedback from supernovae (SNe) can
reduce the baryonic \citep{Dekel1986, Maller2002, Dekel2003b, Brook2011, 
Guedes2011} and DM mass at the centers of galaxies \citep{Navarro1996, 
Read2005, Mashchenko2006, Mashchenko2008, Governato2010, Pasetto2010,
deSouza2011, Cloet2012, Maccio2012, Pontzen2012, Governato2012, Teyssier2012, Ogiya2012}.  
Importantly, there is not yet a unifying baryonic solution that solves the 
missing satellites, cusp/core, and massive subhalo problems simultaneously.

In fact, unlike the missing satellites problem and the core/cusp problem,
solutions to reconcile the predicted massive subhalos with observed 
satellites have not yet addressed the effects of baryons.  For example, 
studies have concluded that the
excess of predicted massive subhalos around Milky Way-mass galaxies
may disappear if satellites are modelled with Einasto (versus NFW)
density profiles, or if the Milky Way's true virial mass is $\sim 8
\times 10^{11}$ M$_{\odot}$, rather than $\sim 10^{12}$ M$_{\odot}$
\citep{DiCintio2012, Vera-Ciro2012,Wang2012}.  It is not yet clear whether these
solutions are themselves sufficient to fully reconcile observations
with CDM-based models, and alternative cosmological models
(e.g., warm or self-interacting dark matter) have also been proposed 
to explain observations \citep{Alam2002, Strigari2007, Lovell2012, Vogelsberger2012, 
vandenAarssen2012}.  

The inclusion of baryonic physics thus remains a glaring gap in predictions for the properties of 
dwarf satellites.  Here we 
investigate whether the inclusion of energetic feedback from stars and supernovae, which has been shown to reduce the central dark matter densities of simulated field 
galaxies, can also reduce the predicted central densities of luminous satellites 
around Milky Way-mass galaxies.  One reason that baryonic effects 
have not been studied in detail in satellite galaxies is because
including baryons in cosmological simulations makes them much more
computationally expensive.  Previous work \citep{Governato2010, 
Guedes2011, Pontzen2012} demonstrated that simply achieving high 
resolution is not enough to reduce the central concentration of dark 
matter in galaxies.  Instead, it is necessary to also adopt a physically 
motivated model in which star formation is limited to high density 
peaks, with densities comparable to the average densities in a giant molecular 
cloud.  This model allows highly over-pressurized regions to form when supernova 
energy is introduced to the high density surrounding gas, and the resulting 
hot bubble of gas flattens the central potential, leading to irreversible 
expansion of dark matter orbits as well as driving a wind \citep{Pontzen2012}.  

The implications of such feedback may be
all the more important when predicting properties of the satellites of
larger galaxies (such as the Milky Way's own dwarf satellite population),
because any effect imprinted on the central densities of the dwarf
galaxies by baryons could be exacerbated during the dwarf galaxies'
tidal evolution around their eventual host. The tidal effects of a disk are expected to be increasingly stronger as the central density profile of satellites becomes shallower \citep{Taylor2001, Stoehr2002, Hayashi2003, Read2006b,Penarrubia2010}. A proper prediction for
both the total number and the internal structure of the Milky Way and
M31's dwarf galaxies may thus require an accurate treatment of
baryonic feedback.  Only a few DM + baryon studies have focused on the
satellite population of a Milky Way-massed galaxy, but at much lower
resolutions than have been achieved in DM-only simulations, making
comparisons with DM-only predictions for inner density profiles
particularly difficult \citep[see, however,][]{DiCintio2012,
 Parry2012}.

In this paper, we compare DM-only and Smooth Particle Hydrodynamics (SPH) simulations 
of two Milky Way-massed galaxies run in a cosmological context to $z=0$.  In fact, 
the simulations used in this work adopt a model that has already been shown to 
alleviate some of the small scale problems of CDM \citep{Governato2010, Oh2011, 
Brook2011, Maccio2012, Pontzen2012, Governato2012}. This model is capable of 
forming bulgeless dwarf disk galaxies that have ``cored'' dark matter density 
profiles \citep[i.e., shallower inner density slopes than predicted by CDM 
simulations without baryons,][]{Governato2010, Governato2012, Oh2011}.  While 
the previous studies have focused on isolated field galaxies, this paper examines 
how this same model affects the satellite population of Milky Way-mass galaxies.

This paper is organized as follows. In Section 2 we describe the simulations 
used in this work, as well as our method of selecting the satellite sample.
In Section 3 we focus on the $z=0$ density and circular velocity profiles of 
SPH satellites, and compare to their DM-only counterparts. Section 4 explores 
the role of supernovae feedback and tidal stripping on the mass dependent 
evolution of satellites, from high redshift to $z=0$.  In Section 5 we propose 
an update to the standard treatment of satellites in DM-only models.
We summarize our results and conclude in Section 6.

\section{Simulations}\label{lm_sims}

The high-resolution N-body $+$ SPH simulations 
used in this paper were run with {\sc gasoline} \citep{Wadsley2004}.  The two 
halos studied in this paper were initially selected from a uniform resolution, 
DM-only, 50 comoving Mpc box.  The initial conditions for this box were generated 
assuming a WMAP Year 3 cosmology \citep{Spergel2007}: $\Omega_m$ = 0.24,
$\Omega_{\Lambda}$ = 0.76, $H_0$ = 73 km s$^{-1}$, $\sigma_8$ = 0.77.  The two 
halos (h277, M$_{vir} = 7\times10^{11} M_{\odot}$, and h258, M$_{vir} = 
8\times10^{11} M_{\odot}$) were selected based on their $z=0$ mass and merging 
histories.  We define virial mass relative to critical density, $\rho_{c}$, using 
$\rho/\rho_{c} = 100$ at $z=0$ following \citet{Gross1997}. Galaxy h277 was selected 
to have a quiescent merger history, with its last major merger occurring at 
$z\sim$3, while h258 has a binary merger at $z=1$. Each of h277 and h258 was 
resimulated at higher resolution (and with gas particles) using the volume 
renormalization technique \citep{Katz1993}.  This approach simulates only the 
region within a few virial radii of the primary halo at the highest resolution,
while still maintaining the large volume at low resolution in order to account 
for the large scale tidal field that builds angular momentum in tidal torque 
theory \citep{Peebles1969,Barnes1987}.  The high resolution simulations were 
run from $z=150$ to $z=0$.

N-body$+$SPH volume renormalized simulations of both h277 and h258 have been 
studied in previously published work.  For example, h258 was the focus of how 
a large disk can regrow by $z=0$ after a low $z$ major merger in 
\citet{Governato2009}.  The results in this paper are based on the same initial
conditions as those previous studies, but simulated at a higher resolution that 
allows for the inclusion of new physics, discussed below.  The spline force 
softening of the high resolution 
regions of both h277 and h258 is 174 pc.  High resolution dark matter particles 
have masses of 1.3$\times$10$^{5}$ $M_{\odot}$, while gas particles start with 
2.7$\times$10$^{4}$ $M_{\odot}$.  Star particles are born with 30\% of the mass 
of their parent gas particle (i.e., a maximum initial mass of 8100 $M_{\odot}$), 
and lose mass through SNe and stellar winds.  Each of these galaxies has roughly 
5 million DM particles within the virial radius at $z$=0, and more than 14 
million particles (dark matter + gas + stars) total.  

The mass and force resolution of these runs is comparable to the ``Eris'' 
simulation \citep{Guedes2011}, one of the highest resolution N-body$+$SPH 
cosmological simulations of a Milky Way-mass galaxy run to date.  
However, unlike Eris, we take advantage 
of the increased simulation fidelity that sufficiently resolves the high 
density regions where stars form ($\rho \sim100$ amu/cm$^3$) to alter the 
gas cooling and star formation prescription to include metal line cooling 
and H$_2$.  The inclusion of metal line cooling allows much more gas to 
cool to the central regions of the galaxy (Christensen et al., in prep.) 
and become fuel for star formation.  However, the star formation in the 
new model is tied directly to the local H$_2$ abundance, which is regulated 
by the gas metallicity and the ability of the gas to self-shield, in accordance 
with observational results \citep{Leroy2008, Bigiel2008, Blanc2009, Bigiel2010, 
Schruba2011}.  The H$_2$ prescription implemented here is described in detail in 
\citet{Christensen2012}.  Briefly, it includes both gas-phase and dust grain 
formation of H$_2$, and destruction of H$_2$ via photodissociation by Lyman-Werner 
radiation from nearby stellar populations.  Before the implementation of the H$_2$ model, gas particles 
were required to be above a set density threshold and below a set temperature 
before they could form stars.  Tying star formation to molecular gas eliminates 
the need to set a density threshold above which stars can form, as it 
ensures that stars form at high densities.  A gas particle must be colder than 
1000K before it can spawn a star particle.

Other aspects of the cooling physics and energy feedback from supernovae 
remain unchanged in these simulations.  \citet{Shen2010} describes the primordial 
and metal-line cooling used, as well as the diffusion of metals that captures 
the effect of the turbulent interstellar medium on metal mixing.  A uniform UV 
background turns on at $z = 9$, mimicking cosmic reionization following a 
modified version of \citet{Haardt2001}.  Star particles are born with a Kroupa 
initial mass function \citep{Kroupa1993}.  Each SN deposits 10$^{51}$ ergs of thermal 
energy within a ``blastwave'' radius calculated following \citet{Ostriker1988}, 
with cooling turned off in the affected region for a time that corresponds to the 
expansion phase of the SN remnant, and described in detail in \citet{Stinson2006}.  
As described in \citet{Stinson2006}, supernova type Ia and II yields are adopted 
from \citet{Thielemann1986} and \citet{Woosley1995}, respectively, and
implemented following \citet{Raiteri1996}.  This prescription was shown to 
reproduce the observed stellar mass -- metallicity relation for galaxies as 
a function of redshift \citep{Brooks2007, Maiolino2008}. 

We demonstrate in this paper that including baryons in these simulations can 
dramatically alter the satellite evolution in comparison to the dark matter only 
case.  Hence, we also have DM-only runs for the two halos described above.  The 
force resolution and $z=$ 0 properties (i.e., number of DM particles and halo 
mass) of these DM-only halos are identical to the SPH runs, though the mass 
of the dark matter particles in the SPH runs is lower by a factor of ($1-f_b$), 
where $f_b$ is the cosmic baryon ratio, $\Omega_b$/$\Omega_m$, and is 0.175  
for the adopted cosmology.  

\subsection{Satellite Selection \& Luminosity Function}

\begin{deluxetable}{lcccc}
\tabletypesize{\scriptsize}
\tablecaption{Satellite Properties}
\tablehead{
\colhead{Satellite ID} &
\colhead{$M_{V}$} &
\colhead{$M_{HI} [M_{\odot}]$} &
\colhead{$M_{HI} [M_{\odot}]$} &
\colhead{$z_{infall}$} \\
\colhead{} &
\colhead{$z=0$}&
\colhead{$z=0$} &
\colhead{ $z_{infall}$}&
\colhead{}  \\
\colhead{ } & 
\colhead{(1)} & 
\colhead{(2)} & 
\colhead{(3)} & 
\colhead{(4)} 
}
\startdata
h258, sat1 & $-13.7$ & 3.5e5 & 1.5e7 &0.6\\
h258, sat2  &  $-12.8$ & 3.4e5 &8.9e6 & 0.9\\
h258, sat3  &  $-12.6$ & 737.5\tablenotemark{a} &6.5e6 & 2.1\\
h258, sat4  &$ -13.3$ & 0. & 3.4e6& 1.3\\
h258, sat5  &  $-13.0$ & 0. &5.7e7 & 2.1\\
h258, sat6 & $ -11.3$ & 1.2e5 & 2.6e6 &1.0\\
h258, sat7 &  $-12.1$ &  0.    & 9.1e5 &1.2\\
h258, sat8 & $-9.0$  &  0.    & 0 & 1.5\\
h258, sat9 & $ -9.9  $& 0.    &2.2e5 &1.8\\
h258, sat10& $ -8.8$  & 0.    & 0 & 1.3\\
h258, sat11 &  $-10.7$ & 0.    &  1.4e6 &1.0\\
h258, sat12 &  $-9.5$  &  0.    &5.4e4 & 2.1\\
h258, sat13 &$ -10.1$ & 0.    & 7.3e3 & 1.2 \\
\hline 
\\
h277, sat1  &  $-14.7$ & 3.2e7 &2.2e7 & 0.02\\
h277, sat2 &  $-13.9$ & 3.8e7 & 3.7e7&0.2\\
h277, sat3  & $-14.3$ & 5.1e6 &3.8e7 & 1.5\\
h277, sat4  &  $-13.4$ & 1.3e6 &3.0e6 & 0.9\\
h277, sat5 & $-10.4$ & 0.    &0 & 1.4\\
h277, sat6 & $-10.6 $&0.    & 9.0e6 & 1.6\\
h277, sat7 & $-9.9  $& 0.    &1.4e5& 1.6\\
\enddata
\tablenotetext{a}{Given the radial distance of this object from the center of 
the main halo that it orbits, this mass of HI would remain undetected by current HI 
observations. We therefore consider it a dSph analog.}
\tablecomments{ Column (1) $V$-band magnitude, calculated based on the age
and metallicity of the star particles and adopting the Starburst99 stellar population
synthesis models of \citet{Leitherer1999} and \citet{Vazquez2005} for a Kroupa IMF
\citep{Kroupa2001}.  Column (2) is the mass in HI 
gas in the satellite at $z$=0.  Column (3)  is the mass in HI 
gas in the satellite at infall.  The redshift of infall is listed in column (4).  }
\end{deluxetable}

Because the main purpose of this work is to study the properties of the 
satellites of Milky Way (MW)-mass galaxies, we explain our method for selecting satellites 
in this section.  Halos and their subhalos are identified at each output 
time step (every 320 million years) using AHF \citep{Knollmann2009,Gill2004}. 
For field galaxies, AHF adopts overdensities with respect to the critical 
density as a function of redshift from \citet{Gross1997}.  At $z$=0 the 
virial radius of h258 and h277 are 240 kpc and 230 kpc, respectively. 

We first find all of the subhalos that are within the $z=$0 virial radii of 
the two primary halos described above.  We then reduce that list to only 
those subhalos that are luminous, with a minimum of 10 star particles. 
This list of subhalos is likely to include some subhalos that have undergone 
a large amount of dark matter stripping, but \citet{Brooks2007} showed that 
the star formation histories (SFHs), and hence stellar masses, of these 
simulations converges only when the halo has more than 
$\sim$3500 DM particles.  Hence, we then trace back the primary progenitor 
halo for each of the $z=0$ subhalos. To do this, we successively identify the
halo at each higher $z$ step that contains the most DM mass of the lower $z$ 
halo.  We verified that all of our $z=0$ luminous satellites have 
had more than 7000 DM particles within their virial radius at some point in the 
halo's history, ensuring that their stellar masses are robust to resolution 
effects.  We exclude bright, Magellanic-like satellites with $M_V < -15$ from this analysis, 
which removes from each simulation one bright satellite.
However, we make no cut on satellite morphology at $z=0$, so that the 
final list of subhalos includes both gas-free and gas-rich satellites.

Table 1 lists some of the properties of the satellites studied in this paper.   
We note that h258 has 13 luminous satellites and h277 has 7 satellites, 
despite their similar halo masses.  This demonstrates  that the number of satellites 
(and particularly the number of highest mass satellites) is very stochastic at a fixed 
parent halo mass \citep{Vera-Ciro2012, Sawala2012}.  

The orbital evolution of every satellite was traced with respect to the parent 
halo.  Although some satellites may enter the virial radius of the parent halo 
and exit again, we identify the infall redshift as the time at which the satellite 
first enters the parent halo's virial radius.  The infall times are listed in 
Table 1, and in all cases the infall redshift is at $z <$ 3 
\citep[see also][]{Zentner2003, Geen2012}.  

\begin{figure}
\includegraphics[angle=0,width=0.5\textwidth]{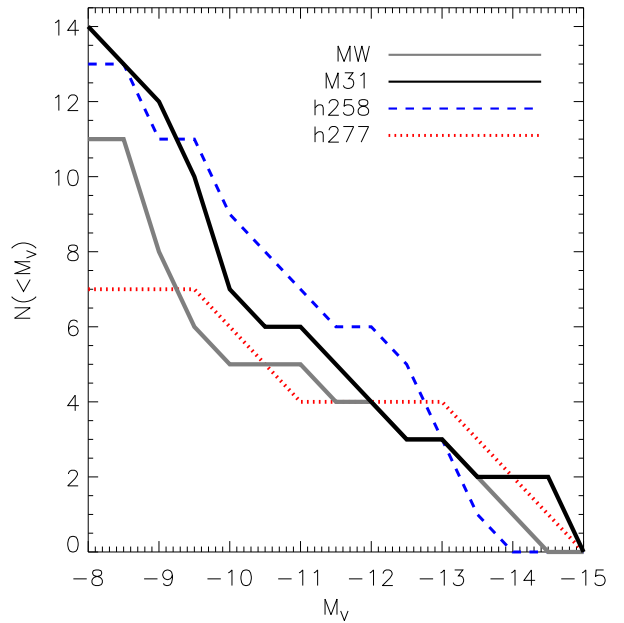}
\caption{The satellite luminosity function for our two simulated galaxies 
compared to the Milky Way and M31 satellite luminosity functions.  The simulated 
luminosity functions contain all satellites listed in Table 1.  }
\label{lf}
\end{figure}

Figure~\ref{lf} shows the satellite luminosity functions for our two 
simulated galaxies compared to that of the Milky Way and M31.  The simulation luminosity 
functions include both gas-free and gas-rich satellites. The MW and M31 luminosity 
functions include all satellites fainter than $M_V = -15$, including Sagittarius and Canis Major in the case of the MW, as compiled in \citet{McConnachie2012}. Our simulated gas-free satellite 
sample spans the luminosity range of the Milky Way's classical dwarf spheroidals, 
$-13.3 < M_V < -8.8$, from Fornax to Canes Venatici I \citep[$M_V = -13.4$ and $-8.6$, 
respectively,][]{McConnachie2012}. 

We stress that no attempt was made to explicitly match the classical dSph 
luminosity range, although our final sample does.   Instead, the deposition of 
SN energy combined with H$_2$-based star formation was implemented to reproduce the 
stellar mass of the {\it parent} halo at the given halo mass.  While past simulations 
have overproduced stellar mass at a given halo mass \citep{Zolotov2009, Guo2010, 
Sawala2011, Brooks2011, Leitner2012}, it has been suggested that restricting 
star formation to depend on the mass in molecular gas (rather than total gas 
mass) will alleviate this problem, particularly at high $z$ where metallicities 
are low and formation of H$_2$ on dust grains is reduced \citep{Robertson2008, 
Gnedin2009, Gnedin2010, Krumholz2011, Feldmann2011, Kuhlen2012}. 
\citet{Munshi2012} demonstrate that the parent halos in this paper 
match the observed $z=0$ stellar mass to halo mass relation \citep{Guo2010, Behroozi2010, 
Moster2010, Conroy2009, Yang2011, Neistein2011, Leauthaud2012}. 
Also, since star formation becomes metallicity 
dependent in this H$_2$-based scheme, it lowers the star formation efficiency 
in the low mass, low metallicity satellites studied in this paper, preventing 
them from overproducing stars.  \citet{Governato2012} previously 
demonstrated that the stellar masses and total masses at $< 1$ kpc for 
simulated field galaxies in the same stellar mass range as the satellite 
galaxies in this paper are in excellent agreement with the observational
data.
 
It is possible that some of the gas-rich satellites included in this work 
may be unphysically gas-rich, because SPH does not accurately model small-scale 
hydrodynamical instabilities that can lead to enhanced gas stripping from 
satellites \citep{Agertz2007, Mitchell2009}.  However, \citet{Weisz2011} have 
recently shown that dSphs with $M_B < -10$ have extended SFHs, suggesting 
that the majority of observed gas-free satellites at $z=0$ had gas until recently. 
Thus, despite the fact that some of our satellites in the luminosity range 
of the MW's dSphs have gas at $z=0$, their SFHs are consistent with those 
observed.  

\subsection{ Matching SPH \& DM-only Satellites}
We have identified the subhalos in the DM-only runs that correspond to each of the luminous satellites in the SPH runs. This matching is performed by requiring that SPH and DM-only satellites have the same properties before infall, as well as similar orbits after infall, in order to be considered a pair. We do this by first compiling a list of candidate matches by identifying the DM-only halos with the best virial mass and position matches at z=3 (prior to infall and any stripping) to the SPH halos. We then identify the infall time and trace the full orbital histories of the candidate halos around the primary galaxy.  In order to be considered a match a DM-only halo must have a similar orbital history around the primary galaxy as the SPH satellite. For three of our satellites, an exact match was not found, and therefore these three satellites are excluded from any direct comparison of matched subhalos in the remainder of the paper.

\section{Redshift 0 Results}

\begin{figure*}
\includegraphics[width=0.33\textwidth]{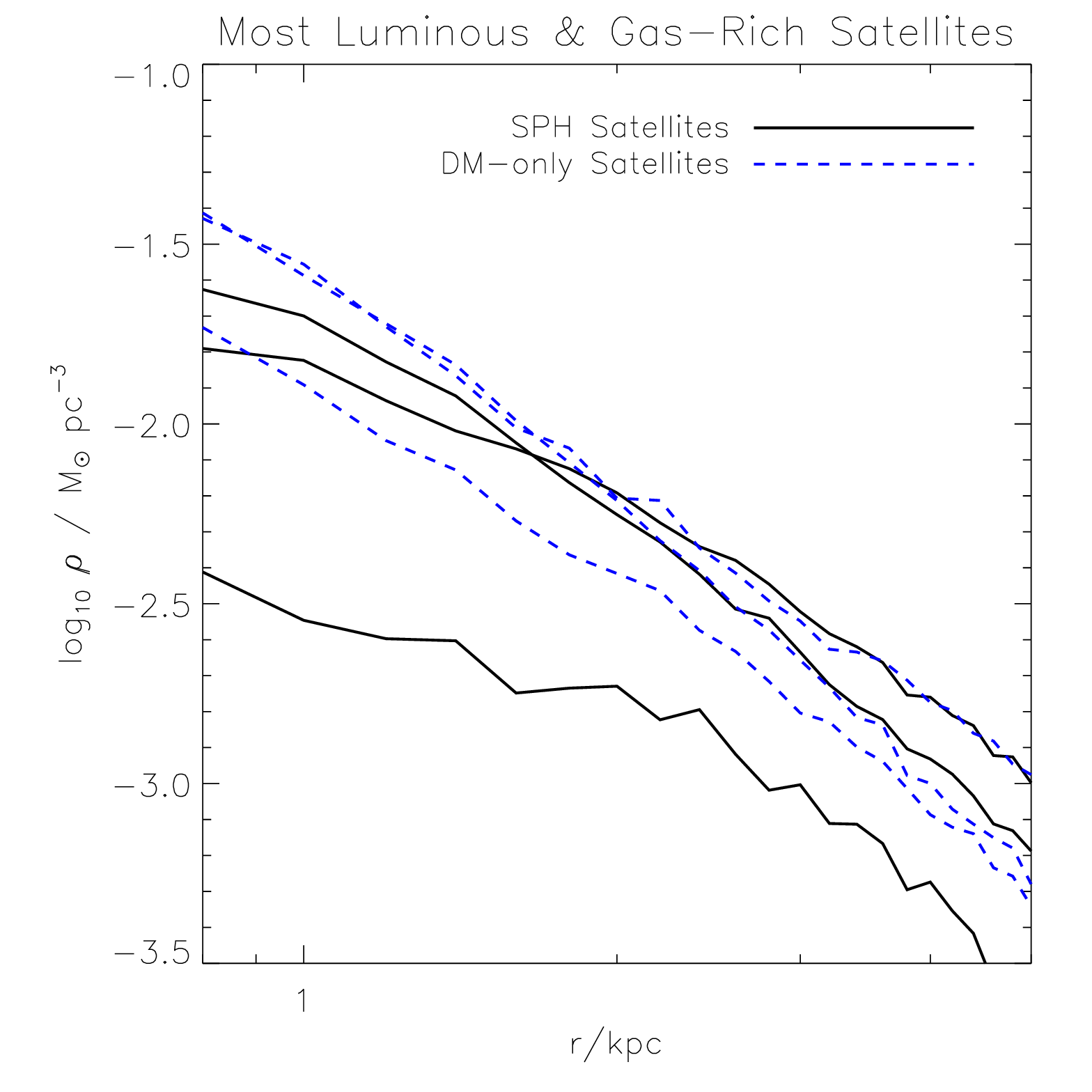} 
\includegraphics[width=0.33\textwidth]{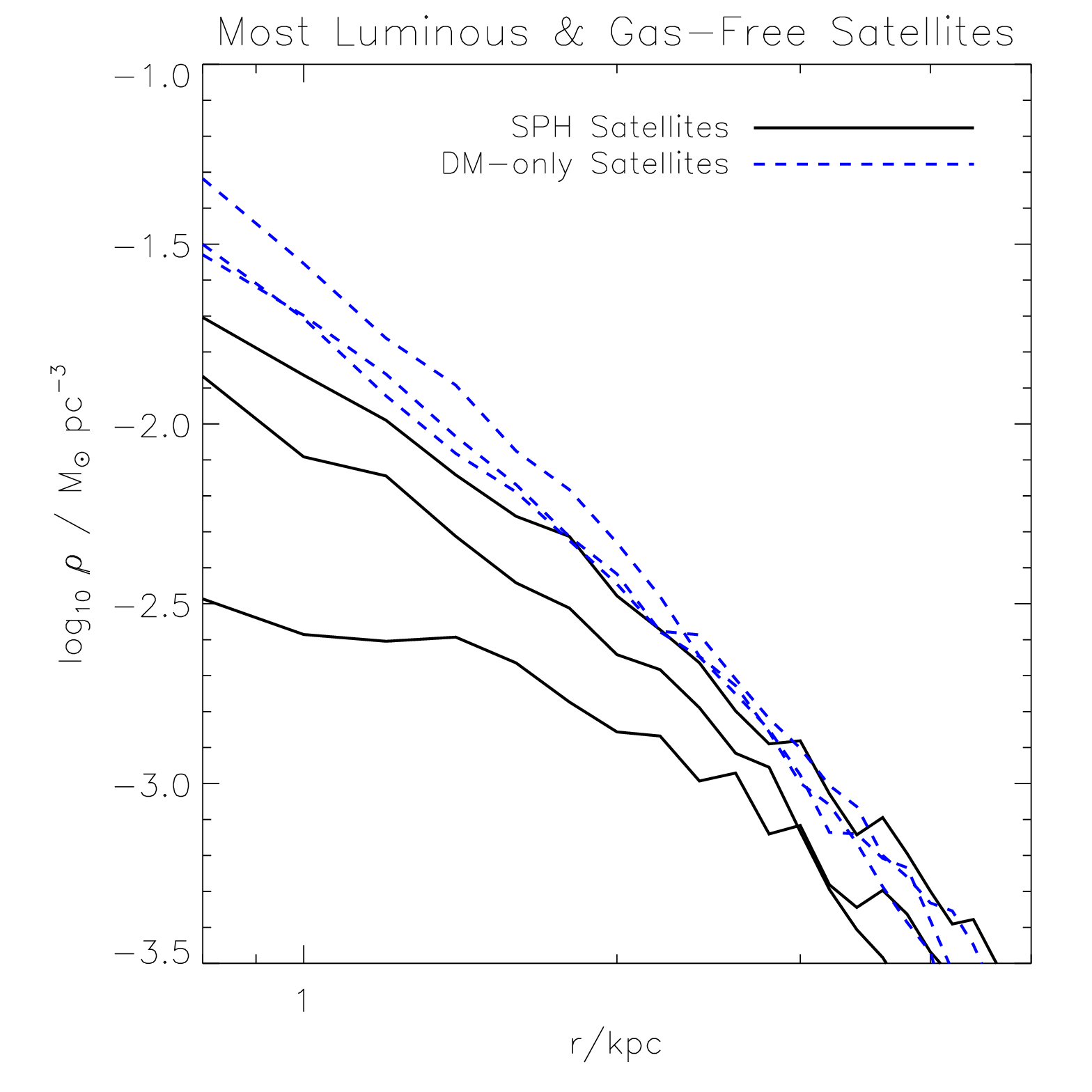} 
\includegraphics[width=0.33\textwidth]{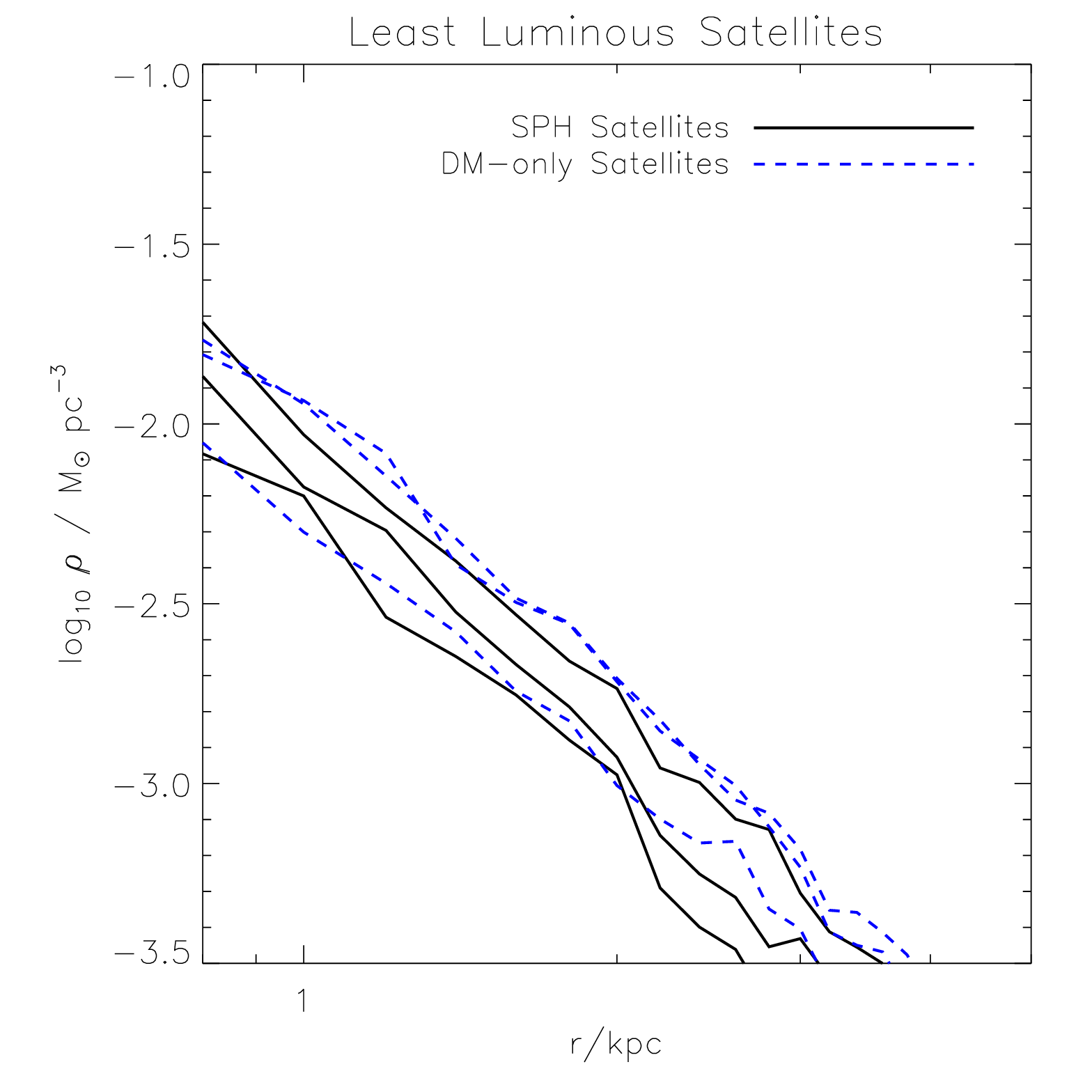} 
\caption{The DM density profiles of SPH satellites (solid black lines) and their 
DM-only counterparts (blue dashed lines) at $z=0$.  The left panel shows the 
three most luminous SPH satellites, which are also still gas-rich at $z=0$. The 
middle panel shows the three most luminous, gas-free satellites, and the right 
panel shows the three least luminous SPH satellites.  
Luminous SPH satellites have significantly shallower central density profiles 
than DM-only satellites, while low-luminosity SPH satellites retain central 
density cusps similar to their DM-only counterparts.}
\label{density}
\end{figure*}

In this Section, we present the DM density profiles of satellites of MW-massed disk galaxies at present day.  In order to study the effect that star formation has had on the internal structure 
of the satellites, we have divided the satellite sample by stellar mass at $z=0$.  ``Luminous'' satellites have 
stellar mass greater than $10^7 M_{\odot}$, while ``low-luminosity'' satellites 
are those with stellar masses less than $10^7 M_{\odot}$. We further subdivide 
the luminous satellites into categories of gas-rich and gas-free. We do this in 
order to distinguish between simulated satellites that would match the MW's dSph 
population, which are all gas-free dwarfs, and simulated satellites that more 
closely resemble the Magellanic clouds, which are gas-rich. All of the satellites 
categorized as ``low luminosity'' are gas-free. 

In Figure~\ref{density} we show the dark matter density profiles, $\rho(r)$, for 
a subsample of the SPH satellites (black solid lines) and their DM-only counterparts 
(blue dashed lines). Each panel in this figure shows satellites from one of the 
three categories we have defined -- the three most luminous {\it gas-rich} satellites 
are shown in the left panel, the three most luminous and {\it gas-free} galaxies in 
the middle panel, and the three least luminous satellites in the right panel.  Figure 
\ref{density} shows that the most luminous and gas rich satellites in the SPH runs 
are less dense  and have flatter inner density profiles than their DM-only counterparts. 
While the luminous gas-free satellites (middle panel) do not appear strongly cored, a 
direct comparison between the SPH and DM-only matches shows that the SPH runs do have 
a more shallow inner profile.  Importantly, these gas-free, luminous satellites have 
dramatically lower densities overall than their DM-only counterparts.  At the lowest 
luminosity end, however, the satellites in the SPH and DM-only runs tend to have 
comparable density profiles (slope and normalization).  Hence, the process that lowers 
the density in our SPH satellites is more effective at the high stellar mass end than 
at the low stellar mass end.  We demonstrate that this process is related to the SFH 
of each satellite in the next section.  We note that because the DM particle masses in the SPH run are 
lower than in the DM-only run by the cosmic baryon factor, $f_{bar}$ (i.e., these 
particles have been split into DM and gas in the initial conditions), we have reduced 
the DM-only densities by $f_{bar}$ in order to make a direct comparison in this figure.

Figure \ref{density} demonstrates that, even at moderate dSph 
galaxy luminosities ($M_V \lesssim -12$, similar to Leo I or And II), baryonic
processes result in an expanded and shallower central dark matter distribution
than predicted by DM-only simulations.  
Even when the
cored\footnote{In the rest of this paper, we will use the term ``cored''
to refer to all slopes shallower than predicted by DM-only simulations
alone.} density profiles of our simulated dwarfs do not have a flat
slope ($\gamma = 0$), the absolute values of density in the central
regions are still dramatically reduced compared to expectations from a
DM-only simulation. We conclude that DM-only simulations, 
or models based on an assumption of adiabatic contraction, make physically 
incorrect predictions for the central ($<$ 1 kpc) masses of dwarf galaxies 
more luminous than Leo I ($M_V = -11.9$).  Such models should therefore be used 
with caution when used to interpret the observed dynamics
of dwarf satellites in a cosmological context.  

The general conclusions discussed in this paper are independent of resolution effects, as is discussed in detail in the Appendix.
In the remainder of the paper we will often compare the circular velocity, $v_c$, values in the SPH and DM runs at 1 kpc.  As will be seen in the next section, the $v_c$ values at 1 kpc show 
the dramatic effect that baryonic physics has on the SPH runs, but avoids 
biasing this value due to convergence issues at smaller radii.  
More importantly, the $v_c$ values at 1 kpc between the SPH and DM-only runs 
are in excellent agreement in our low-mass subhalos for which baryons do 
not dramatically alter the evolution. The convergence of SPH and DM-only 
results in these lower mass halos (which, by definition, contain fewer 
particles and are less resolved than their high mass 
counterparts) demonstrates that there are no spurious numerical effects 
introduced by the lower mass baryonic particles in the SPH runs.

\section{Mass Dependent Evolution of Satellites with Baryons}
In this Section, we study the evolution of the satellites 
to understand  the processes that lead to the lower concentration of mass at $z=$0 in the 
SPH satellites.  We first focus on the evolution at high redshift, and demonstrate that DM core creation occurs in the most luminous satellites prior to their infall. After infall, we show that tidal stripping effects exacerbate the mass discrepancy between SPH satellites and their DM-only counterparts.

\subsection{The Impact of Baryons Before Infall}

\begin{figure}
\includegraphics[angle=0,width=0.5\textwidth]{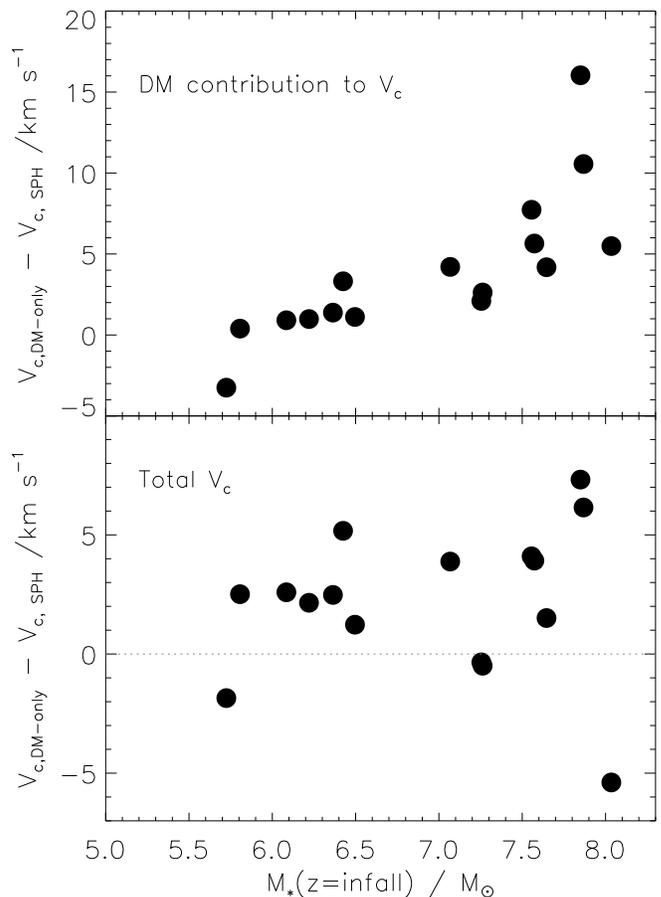}
\caption{The difference between $v_c$ at 1 kpc in the SPH and DM-only runs 
at infall, as a function of the stellar mass in the SPH satellite at 
infall.  {\it Top panel:} The difference in the DM contribution to $v_c$ 
at 1 kpc for matched SPH and DM-only subhalos. 
{\it Bottom panel:} The difference in total $v_c$ at 1kpc.  }
\label{smass}
\end{figure}

\begin{figure*}
\includegraphics[width=0.3\textwidth]{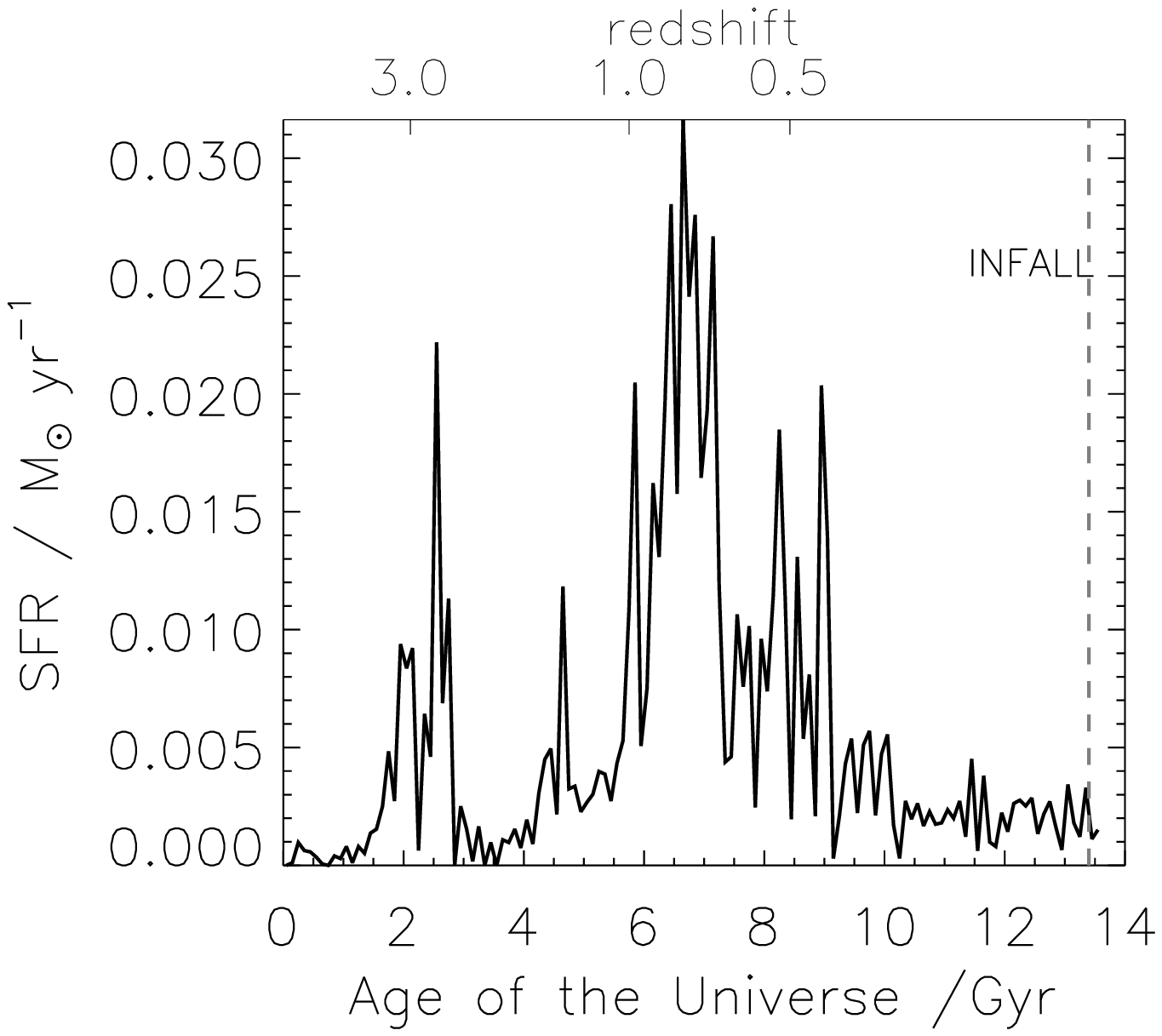}
\includegraphics[width=0.3\textwidth]{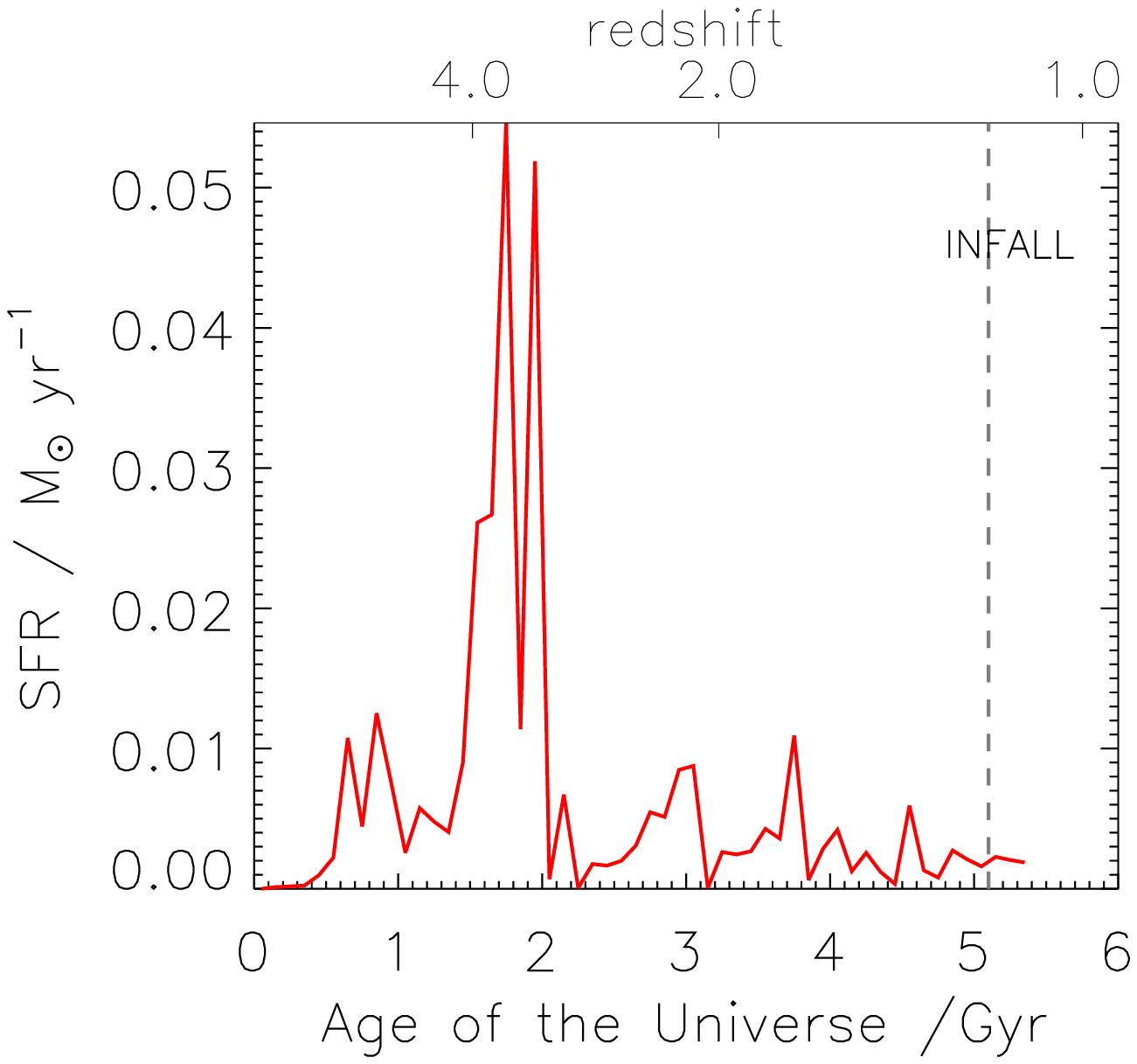}
\includegraphics[width=0.3\textwidth]{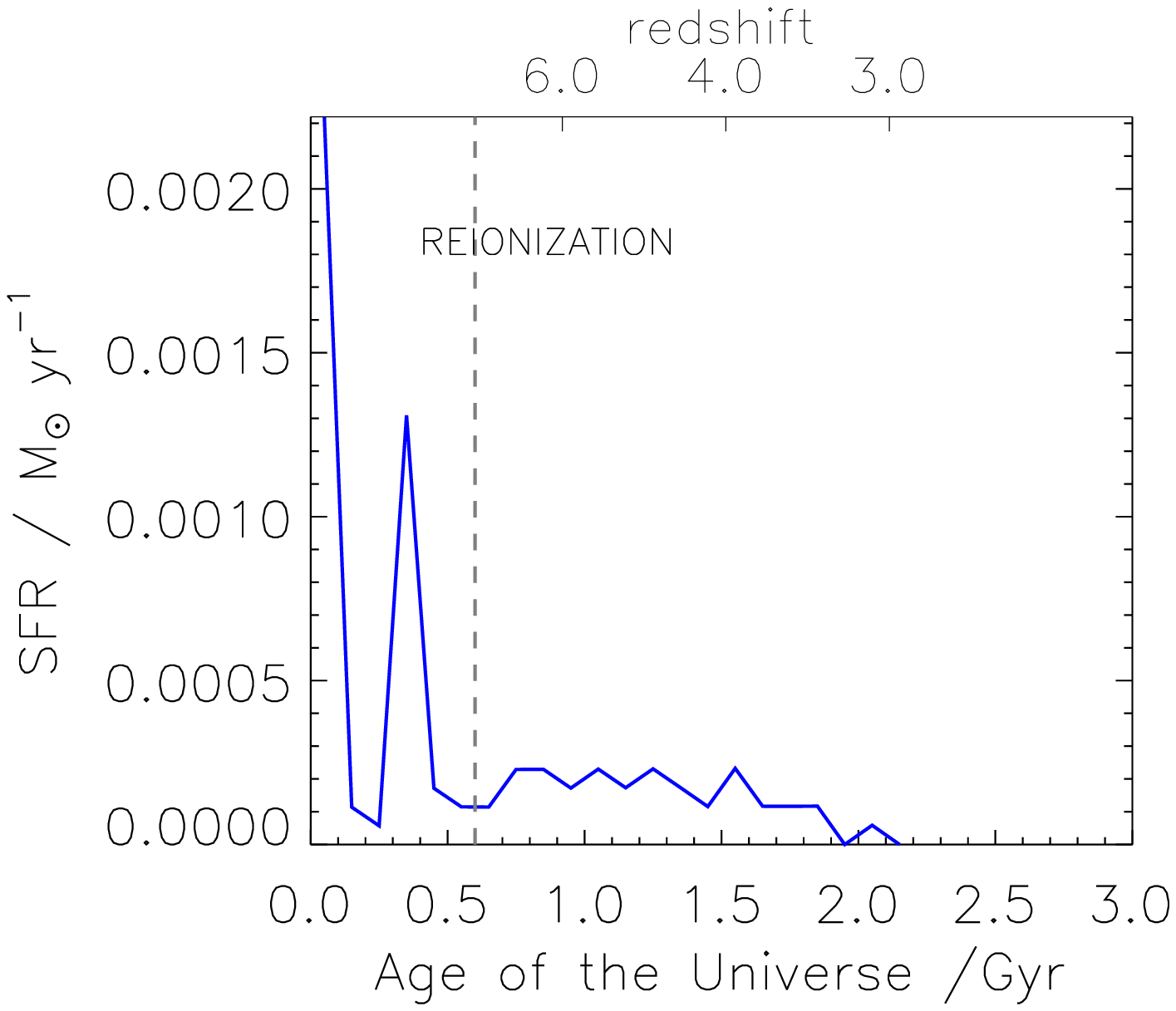}
\caption{The star formation histories of three SPH satellites. 
Note the different range for the axes. At infall, the satellites in black 
(left panel), red (middle panel), and blue (right panel) have virial masses of 
$7.9 \times 10^9$, $2.7 \times 10^9$, $0.6 \times 10^9$ $M_{\odot}$, respectively, and 
infall times of $z=0.02, 1.3, 1.5$. In the first two panels, the vertical 
dashed line marks $z=$infall onto the 
parent halo.  In the final panel, $z=$infall lies 
beyond the range of the x-axis, which is truncated before infall for clarity.  
However, the vertical dashed line in the right panel indicates the onset 
of reionization (i.e., when the uniform UV background turns on). }
\label{sfh}
\end{figure*}

We now examine the evolution of $v_c$ from high $z$ to infall for all of 
the satellites in our sample.  The top panel of Figure~\ref{smass} shows the 
change in the DM contribution to $v_c$ at 1 kpc between SPH satellites and 
their DM-only counterparts at infall, as a function of the stellar mass of 
the SPH satellites at infall. 
It can be seen in this panel that satellites that have formed more stars prior to infall (those 
with M$_{\star} > 10^7 M_{\odot}$ at infall) undergo a significant decrease 
in DM mass interior to 1 kpc, in comparison to their matched counterparts 
in the DM-only runs. The DM rotation curves of these luminous satellites are 
therefore  $2-16$ km/s lower than those of DM-only satellites. (Masses in the 
DM-only run have been reduced by $f_{bar}$ for a direct comparison.) We 
conclude that baryonic effects lower the central DM densities, and hence lower 
the central DM circular velocity, of massive satellites {\it prior} to 
infall.  

The bottom panel of Figure~\ref{smass} shows the total change in $v_c$ at 1 kpc 
between SPH satellites and their DM-only counterparts at infall. 
This panel shows that the overall reduction in total $v_c$ is 
is not as strong as the reduction in DM $v_c$. Although the DM mass has been 
reduced for satellites with M$_{\star} > 10^7 M_{\odot}$, the presence of baryons contributes 
to the central masses in such a way that the overall mass is not necessarily 
reduced compared to the DM-only case prior to infall.
 Note that this is not adiabatic contraction, in which the central DM densities are increased due to the cooling of baryons, as we have just demonstrated that the baryonic runs have comparable or lower DM densities to the DM-only runs. 
We demonstrate below that once gas is stripped from the SPH satellites after infall, the reduction seen in 
the DM masses and the reduction in total mass are in agreement.  The global 
trend of $\sim$3 km/s reduction in total $v_c$ in the SPH runs is due to gas lost in either SNe-driven outflows or reionization.  

\citet{Governato2012} also found that DM core creation varied as a function of mass, 
for isolated field galaxies.  These authors have shown that the creation of DM 
density cores due to SNe driven outflows is common in galaxies with 
$M_{\star} > 10^7$ at $z=0$.  For lower luminosity galaxies, their work 
finds that SF and its associated feedback are not efficient enough to have flattened 
a galaxy's steep DM density profile.  We find that this mass threshold above which
feedback becomes effective is the same in satellite galaxies as well, despite the 
fact that core formation does not continue after infall. 

Several theoretical models have shown how outflows and galactic fountains can 
lead to the flattening of DM density cores in dwarf galaxies \citep{Navarro1996, 
Read2005, Pontzen2012, Governato2012}.  Rapid and frequent star formation episodes 
break the adiabatic approximation in the central kiloparsec of galaxies, 
transferring energy to the collisionless particles, and resulting in shallow 
DM density cores \citep{Pontzen2012, Ogiya2012}. An essential component to such 
DM-flattening scenarios is the ability to resolve the high density clumps where 
star formation takes place, i.e., overdensities comparable to giant 
molecular clouds \citep{Robertson2008, Saitoh2008, Tasker2008, Ceverino2009, 
Christensen2010, Colin2010}.  When star formation is limited to these 
high density peaks, energy deposited from SNe creates overpressurized 
regions of hot gas, driving outflows of SNe heated gas from the 
simulated galaxies. 
The H$_2$ model adopted in these simulations includes self-shielding 
of cold gas, preventing heating from photoionization in dense regions with no 
young stars nearby.  Self-shielding, combined with a low temperature cooling model, 
allows the already cool gas in the galaxy to cool even further, making the star 
formation even clumpier and the feedback therefore even more efficient 
\citep[][Christensen et al., in prep.]{Christensen2012, Susa2008}. 

When star formation is limited to high density peaks, as in this paper, the 
SNe feedback after a star forming event creates regions of overpressurized, hot 
gas that will shut down additional star formation for a period of time until 
the gas can cool and continue to form more stars.  This process leads to a 
bursty SFH.  Examples of the resulting star formation histories for galaxies 
in this paper are shown in Figure~\ref{sfh}.  Dwarf galaxies in our highest luminosity range are massive enough to retain gas for
an extended period of time, allowing them to undergo multiple star forming
and gas loss events. Each event pushes successively more DM to larger orbits, 
gradually transforming a cuspy DM density profile into a flatter cored profile.
Hence, extended SFHs that allow for multiple bursts of SF lead to effective DM 
core creation.  At the lower end of the luminosity range of our satellites, the 
halos are less massive, with shallower potential wells.  A few are low enough 
in mass that they lose their gas relatively early, though most maintain some gas 
at least until infall.  The overall gas mass in these low mass subhalos is 
substantially reduced at early times, due to a combination of reionization and 
initial SF that ejects gas \citep[see also][]{Sawala2010}.  The remaining gas in 
the shallow potential well then has a difficult time reaching the 
densities required for star formation.  The inability to continue forming stars 
in the lower mass, less luminous satellites prevents them from having bursts of 
star formation over extended periods, and they thus retain a steeper DM density 
profile.  

\begin{figure}
\includegraphics[angle=0,width=0.5\textwidth]{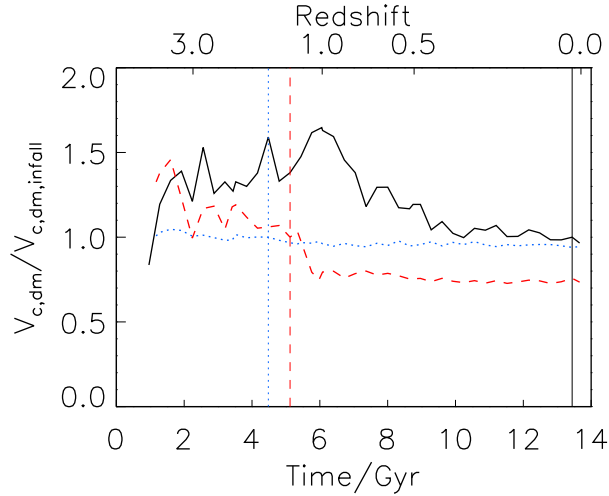}
\caption{The DM contribution to $v_c$ at 1 kpc in the three SPH satellites shown 
in Figure~\ref{sfh} as a function of time.  The $v_c$ value 
at all times is normalized to the value at infall. A luminous gaseous (at $z=$0) 
satellite is in black (solid line), a luminous gas-free (at z=0) satellite is in red 
(dashed line), and a low luminosity gas-free (at z=0) satellite in blue (dotted line).  
Their respective infall times are shown by the vertical lines, with colors and 
linestyles according to galaxy as just described.  In the two most luminous 
satellites, decreases in $v_c$ correspond to bursts of SF seen in Figure~\ref{sfh}. }
\label{vcevol}
\end{figure}

If the mass trends described above are accurate, we should see these trends 
reflected concurrently in the SFHs of the galaxies and their reduction in 
central dark matter mass. To examine the evolution of the satellites as a 
function of mass, we focus on the progenitors of three satellites as examples, drawn from 
the three different categories represented in Figure~\ref{density} (gas-rich, 
gas-free and most luminous, gas-free and least luminous). Figure~\ref{sfh} 
shows the SFHs of these three representative satellites. The SFHs, 
particularly in the luminous satellites, are episodic with bursts, rather 
than smooth or constant. In Figure~\ref{vcevol}, we show the evolution of the 
DM contribution to $v_c$ at 1 kpc in these same three galaxies. 
Each $v_c$ evolution curve is 
normalized to the $v_c$ value of the DM at infall, where the infall times are 
marked by vertical lines, color coded to each galaxy. 

For the luminous satellites (shown in black and red in Figures~\ref{sfh}
and \ref{vcevol}), reductions in the DM contribution to $v_c$ at 1 kpc can be seen 
following strong bursts of SF in these galaxies. For example, a DM core begins to 
form in the progenitor of the most luminous satellite (shown in black) near $z=1$, 
consistent with a large number of repeated bursts of SF that begin at this time.
Figure~\ref{vcevol} indeed shows a significant decrease in the central DM mass 
associated with core formation beginning at $z \sim 1$. During periods of low SF, 
baryons have little to no impact on the DM structure of these satellites.  For 
example, the satellite shown in red does not undergo any changes in the 
contribution of DM to its central $v_c$ between $z=3$ and infall, which is tied 
to the weak SF of this galaxy during this period.

The low luminosity, gas-free satellite (shown in blue in Figures~\ref{sfh} and 
\ref{vcevol}) forms most of its stars prior to reionization. The gas supply of this 
satellite is strongly affected by reionization, with SF declining in the $\sim3$ Gyr 
afterward, until gas is completely unbound from this halo at $z \sim2$. The single, 
small burst of SF in this galaxy prior to reionization was not enough to alter the DM 
density profile, nor was the steady decline in SF between reionization and $z=3$. 
This is evidenced in Figure~\ref{vcevol}, which shows that the DM contribution 
to $v_c$ in the central 1 kpc stays constant through the entire history of this 
galaxy, both prior to and after infall.

Finally, we verified that the SFHs of satellites are either completely truncated 
at infall, or SF continues at a low rate with no strong bursts of SF that are 
associated with core formation.  Thus, we conclude that the dark matter core 
formation occurs prior to infall.  The SFHs of the galaxies before infall are 
intimately tied to their DM density profiles, with multiple SF bursts required 
to create a DM core in the progenitor of a satellite \citep{Read2005, Pontzen2012}.

\subsection{Evolution After Infall}

In this Section, we focus on the evolution of the central mass of satellites after infall in order to explore whether the presence of baryons in galaxies
leads to diverging tidal evolution between the SPH and DM-only cases.  In tracing the orbits of our satellite sample, we found that 
some satellites undergo significant mass loss after accretion, with most 
of the mass loss occurring at pericenter passage where tidal effects are 
strongest.  This is true in both the SPH and DM-only runs, a result that 
has been seen in many other studies \citep[e.g.,][]{Mayer2001, Dekel2003b, 
Hayashi2003, Kazantzidis2004, Mayer2006, Read2006, Boylan-kolchin2007, 
Maccio2009, Choi2009, Klimentowski2010, Kazantzidis2011}.  However, most work 
to date on mass loss in satellites has  
excluded the effects of baryons. We demonstrate 
below that when a satellite's orbit is such that it is likely to 
undergo significant tidal stripping, SPH satellites will lose a substantially larger fraction of their central mass after infall compared to their DM-only counterparts.

\begin{figure}
\includegraphics[angle=0,width=0.5\textwidth]{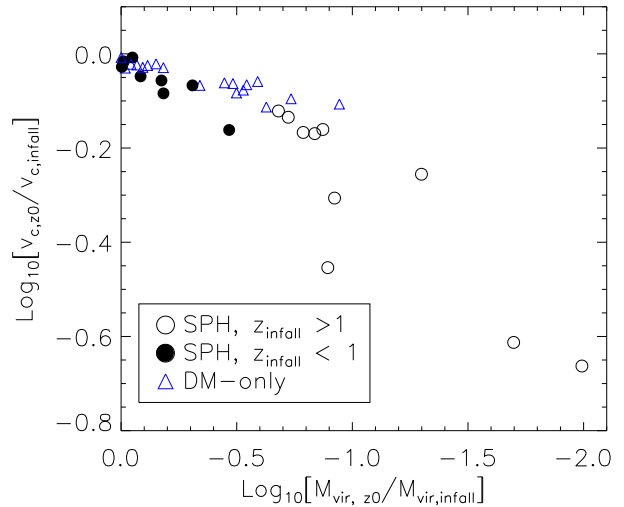}
\includegraphics[angle=0,width=0.5\textwidth]{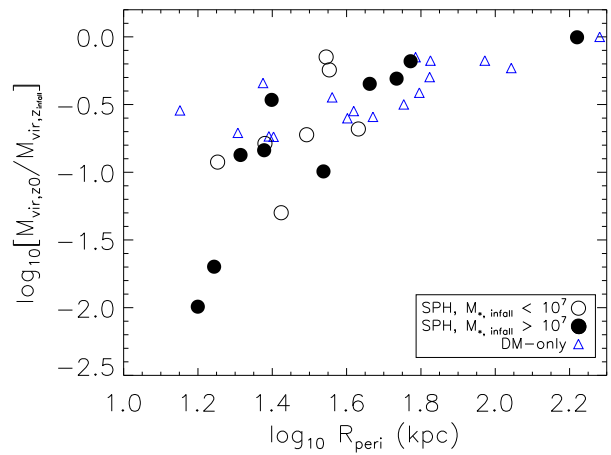}
\caption{Top Panel: The change in $v_{c}$ at 1 kpc between $z=$infall and $z=$0 for the simulated satellites, as a function of the fraction of mass retained since infall.
The SPH satellites (in circles) are divided by their infall time, and the matched DM-only counterparts are shown as blue diamonds. Bottom Panel: The fraction of mass retained after infall by satellites, as a function of the minimum pericenter radius in their orbital history. The SPH satellites (in circles) are divided by their stellar mass at infall time in this panel.}
\label{pena}
\end{figure}

In the top panel of Figure~\ref{pena} we show the change in circular velocity, between infall and $z=0$, at 1 kpc 
for all simulated satellites, as a function of the fraction of mass retained since infall. Two important trends emerge in this figure. First, SPH satellites tend to undergo a larger decrease
in their $v_c$(1 kpc) than their DM-only counterparts after infall. We find that the relative reduction of $v_c$ between SPH and DM-only satellites is highly dependent on the orbital parameters of the satellites. For example, satellites accreted more than 6 Gyr ago ($z_{infall} > 1$) 
experience a reduction in their central circular velocities that is $11 - 62 \%$ more than the reduction in $v_c$(1 kpc) experienced by their DM-only counterparts.  Likewise, satellites on more radial orbits ($r_{peri}:r_{apo}> 8$) undergo a reduction in $v_c$ that is $35\%$ greater than their DM-only matches. On the other hand, SPH satellites accreted more recently ($z_{infall} < 1$) or on circular orbits ($r_{peri}:r_{apo}< 2$) have their $v_c$ reduced by only $0-13 \%$ more than their DM-only counterparts.  A larger decrease in $v_c$(1kpc) will occur in some of the SPH satellites simply from gas loss (e.g.,  in ram pressure stripping).  A comparison of the HI masses at infall to the $z=0$ HI masses in Table 1 demonstrates that significant amounts of gas are lost.  In the vast majority of the satellites, however, the decrease in $v_c$ in the SPH satellite is significantly higher than can be accounted for by just loss of gas. The second trend seen in the top panel of Figure~\ref{pena} is that only SPH satellites lose 
more than than $90\%$ of their total mass.\footnote{Subhalos must have lost $\sim$90\% of their mass before 
stars, which are more tightly bound, begin to be stripped \citep{Penarrubia2008}. 
Only three satellites in our sample change their stellar mass by more than 20\% 
between infall and $z=$0, and they are the same three that 
have lost more than 90\% of their total mass.}  Overall, we find that SPH satellites lost $3 - 34\%$ more of their virial mass than their DM-only counterparts after infall.

\citet[][hereafter P10]{Penarrubia2010} used N-body simulations to examine the role of different DM density profiles, as well as the role of a baryonic disk in the host, on the tidal evolution of satellites \citep[see also][]{Taylor2001, Stoehr2002, Hayashi2003,Kazantzidis2004,Read2006b,Choi2009,D'Onghia2010,Romano2010,Wetzel2010,Nickerson2011}. Their models show that the presence of a baryonic disk in the primary galaxy results in a higher mass-loss rate for satellites at each pericentric passage, in comparison to a host with no disk.  The influence of a baryonic disk is especially strong for satellites with shallow DM density profiles, with slopes of  $0.0 \le \gamma \le 0.5$, resulting in a higher fraction of mass loss at each pericenter passage for the cored satellites. The combined effects of the host disk and a satellite's shallow DM density profiles in P10 becomes increasingly significant with increasing time after infall and for satellites on more eccentric orbits. 

The greater reduction in central circular velocities and virial masses we find in our SPH satellites is due in part to the presence of a disk in the SPH host galaxies (the DM-only runs do not have a disk), which results in more efficient mass loss for the SPH satellites. Indeed, the three SPH satellites that have lost the most mass ($> 90 \%$), are all on orbits whose pericenters bring them within the inner $30$ kpc of the galaxy, where the effects of the disk on mass loss are strongest. This can be seen in the bottom panel of Figure~\ref{pena}. Furthermore, the two satellites that lost the most amount of mass both had shallow DM density profiles at infall (with $M_{*} > 10^7 M_{\odot}$ at infall). As noted in P10, the effect of the disk should be more pronounced in the tidal evolution of cored satellites, since the disk can dominate the tidal field in the center of cored satellites, but not in the tightly bound cuspy halos. Indeed, our samples contains two SPH satellites -- one with a cored DM density profile and one with a cuspy DM density profile -- that have similar infall times, orbital histories, and both have $R_{peri} < 30$ kpc. The $v_c$ at 1 kpc of the cored SPH satellite was reduced by $78\%$ after infall, while the cuspy SPH satellite underwent only a $44\%$ reduction in its central $v_c$. 

We conclude that {\it if} a satellite is likely to have undergone significant 
tidal stripping due to its orbital history, {\it the existence of a baryonic disk in the host and a shallow 
inner DM density profile in the satellite will exacerbate the amount of mass lost.}

\section{An Update to the Theoretical Model}

\begin{figure}
\includegraphics[angle=0,width=0.5\textwidth]{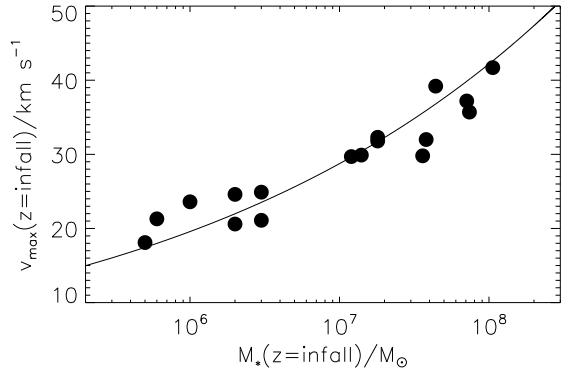}
\caption{The $v_{max}$ values of the SPH subhalos at infall as a function of their stellar mass at infall. The solid line shows the relation M$_* \propto$ $v_{max}^6$, normalized at $M_* =$ 10$^7$ M$_{\odot}$, indicating how stellar mass increases with $v_{max}$ in these simulations.}
\label{sm2hm}
\end{figure}

When using DM-only theoretical results to interpret the MW's dwarf satellite 
population, it is common to associate the DM-only subhalos that were the most 
massive at the time of their formation or accretion with the more luminous 
dSphs \citep{Bullock2000, Kravtsov2004, Gnedin2006, Strigari2007b,
Koposov2009, Bovill2011, Simha2012}, i.e., an abundance matching technique.  This assignment 
is reasonable if we assume that the most massive halos will have the highest star 
formation rates, and therefore be the most luminous 
at $z=0$.  Figure~\ref{sm2hm} shows that our simulated halos follow a trend of 
increasing stellar mass with halo mass (represented by $v_{max}$) prior to 
infall, so that the most luminous galaxies at $z=0$ indeed correspond to the most 
massive galaxies at infall. We verified that the matched DM-only counterparts 
can be assigned to the most luminous satellites at infall in the SPH run.
In these simulations, M$_* \propto$ M$_{vir}^2$ 
\citep[see][]{Governato2012}.  Because $v_{max}$ roughly scales as M$_{vir}^{1/3}$ 
\citep[e.g.,][]{Klypin2011}, M$_* \propto$ $v_{max}^6$.  This relation is shown as 
the solid line in Figure~\ref{sm2hm}, normalized at M$_* =$ 10$^7$ M$_{\odot}$.
We note, however, that tidal stripping can substantially reduce the mass
of some halos by $z=0$, introducing scatter into the tight relation seen in 
Figure~\ref{sm2hm} at infall.

\begin{figure}
\includegraphics[angle=0,width=0.5\textwidth]{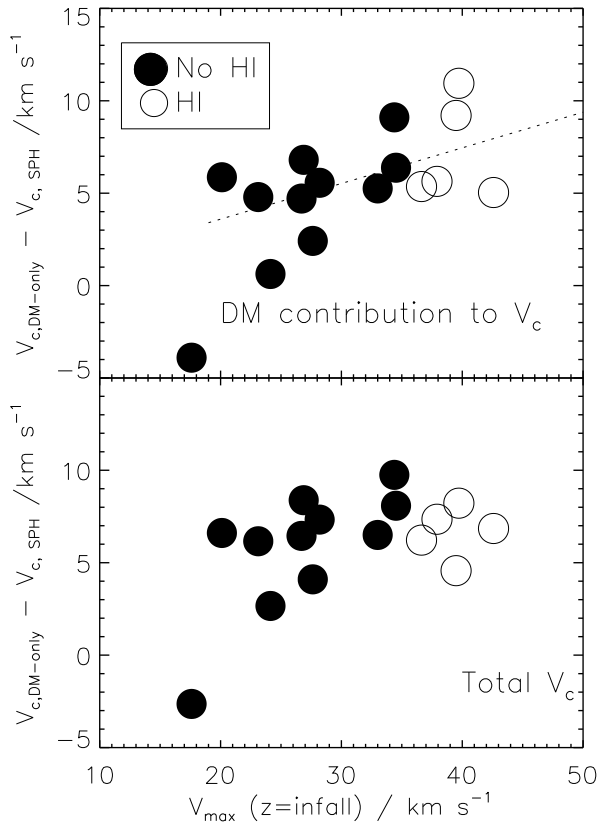}
\caption{The difference in $v_c$ at 1 kpc at $z=0$ between the SPH and 
DM-only counterparts, as a function of $V_{max}$ of the DM-only satellite 
at infall.  {\it Top panel:} The difference in the DM contribution to $v_c$ 
at 1 kpc for matched SPH and DM-only subhalos.  {\it Bottom panel:} The 
difference in total $v_c$ at 1kpc.  }
\label{summary}
\end{figure}

While the assumption that stellar mass increases with halo mass is valid 
for these simulations, it cannot be assumed that the {\it central} dark matter mass 
distribution of the DM-only subhalos is the same as that of luminous subhalos. 
Despite the fact that the virial masses at infall are similar in the SPH and 
DM-only runs, the inner DM orbits have been expanded in the SPH runs (but, prior to 
infall, are not removed from the halo).  As this paper has shown, the DM mass in 
the inner regions is lowered for those satellites with M$_* > 10^7 M_{\odot}$ due to the effects of 
feedback prior to infall.  After infall, SPH satellites are also prone to significantly more 
tidal stripping than their DM-only counterparts.  Figure~\ref{summary} 
summarizes our results on the combined impact of DM core-creation and tidal 
stripping on the internal dynamics of satellites. This figure shows the difference
in $v_c$ at 1 kpc at $z=0$ between SPH and DM-only satellites, as a function of $v_{max}$ at 
infall of the DM-only satellites. The top panel of this Figure shows the change in 
the DM contribution to $v_c$ (again, the DM-only masses have been reduced by $f_{bar}$ 
for a direct comparison), while the bottom panel shows the change in the total $v_c$ 
at 1 kpc.  We note that we exclude from the plot one data point that has a change in 
total $v_c \sim25$ km/s.  This dramatic 
change is due to the fact that the SPH satellite has recently passed directly 
through the disk of its parent galaxy, stripping it more substantially than 
its DM-only counterpart (since there is no disk in the DM-only run).

A well defined trend for the change in DM mass interior 
to 1 kpc between SPH and DM-only runs can be seen in the top panel of Figure~\ref{summary}.
In order to quantify the change in DM mass that baryonic processes combined with 
tidal stripping induce in the central regions of satellites compared to the 
DM-only case, we fit a linear regression to the data points in the range $20 
< v_{max} < 50$ km/s. The 
resulting fit, shown in the top panel of Figure~\ref{summary} as the dashed line, is 
$\Delta(v_c,1kpc) = 0.2 v_{max, DM-only}-0.26$ km/s.  

The bottom panel of Figure~\ref{summary} shows that the change in total 
$v_c$ at 1 kpc between SPH and DM-only runs is slightly lower for gaseous 
satellites (open circles) than those satellites that are gas-free.  Like the 
bottom panel of Figure~\ref{smass}, this is because baryons contribute significantly 
to the mass in the interior 1kpc.  While there is not much trend at infall visible 
in the bottom panel of Figure~\ref{smass}, a trend is beginning to emerge at $z=0$ 
in the bottom panel of Figure~\ref{summary}.  That is, as gas is removed from these 
systems after infall, the underlying difference in the DM central masses in the 
SPH and DM-only runs emerges.  Note that it is likely that the gas-rich satellites in this work 
are artificially gas rich due to inefficient stripping of gas in SPH simulations. The
trend in the bottom panel of Figure~\ref{summary}, therefore, would be even stronger if SPH did not suffer from this numerical effect.

We advocate that DM-only results apply the above linear relation to those 
halos that are accreted onto Milky Way-mass halos with $20 < v_{max} < 50$ km/s.  
Failure to account for these effects will over predict the central mass of 
luminous satellites such as those around the Milky Way and M31.
This is likely the source of the tension discussed in 
\citet{Boylan-kolchin2011,Boylan-kolchin2012} between 
the densities in the MW's classical dSph population and the DM-only 
Aquarius subhalos.  Those studies emphasized that the observed MW 
dSphs appear to be less dense than those found in a CDM DM-only simulation.
Including the affects of baryonic physics reduces, and may even 
completely alleviate, this discrepancy.  We explore the observational 
consequences of our updated model further in a companion paper 
(Brooks \& Zolotov 2012).

\section{Conclusion \& Discussion}

We have demonstrated that the inclusion of baryonic physics can dramatically 
alter the evolution of bright satellite galaxies around Milky Way-massed 
galaxies.  
The population of satellites studied here have $V$-band magnitudes 
consistent with the range of the Milky Way's classical dSphs,
$-15< M_V < -8$. 
Our sample contains both gas-free dSph analogs and massive gas-rich dwarfs. 
By directly comparing the internal properties of satellites 
simulated with gas hydrodynamics (SPH) to the same satellites in 
DM-only simulations, we have demonstrated the impact of baryons 
across a range of masses on satellite evolution. Our main results are summarized below.

\begin{itemize}
{\item Before infall, the progenitors of luminous satellites ($M_{vir} > 10^9 M_{\odot}, M_{\star} > 10^7 M_{\odot}$) undergo rapid and frequent bursts of star formation. The associated SNe feedback from these SF episodes results in structural changes to the central mass distribution of these dwarf galaxies, resulting in reduced DM densities and shallower inner DM density profiles than DM-only galaxies. } 

{\item The progenitors of lower luminosity satellites ($M_{\star} \le 10^7 M_{\odot}$ 
at infall) reduce their gas content, and hence become less efficient at forming 
stars, earlier than more massive dwarfs. This gas loss is partly due to heating by 
the uniform UV background, as well as subsequent gas loss in early star 
forming/feedback events, preventing them from having multiple strong bursts of star 
formation that lead to DM core creation. Low luminosity satellites, therefore, tend 
to retain steep DM density profiles that are comparable to DM-only runs. }

{\item For SPH satellites across all masses, the overall reduction prior to infall in total $v_c$ is, on average, less than 5 km/s.  Although SNe-driven outflows have reduced the central DM mass in halos with $M_{vir} > 10^9 M_{\odot}$, the presence of gas keeps the overall central mass comparable to the DM-only case.  However, this gas is stripped after infall.  Hence, the major reduction in central mass is set in place within the DM component prior to infall, but the removal of gas is necessary to reduce the total mass between the SPH and DM-only runs by z=0.}

{\item Once accreted, SPH satellites experience more mass loss due to tidal stripping than DM-only satellites, the amount of which is dependent on infall time and orbit.  While both SPH and DM-only satellites are affected by tidal stripping, the presence of a baryonic disk in the SPH runs results in a greater reduction in the central $v_c$ in SPH satellites. The influence of a baryonic disk is especially strong for satellites with shallow DM density profiles. We find that SPH satellites with $z_{infall} > 1$ experience a reduction in their central circular velocities that is $11 - 62\%$ more than the reduction in $v_c$ at 1 kpc experienced by their DM-only counterparts. }

{\item By $z=0$, the combined effects of DM core creation and enhanced tidal stripping for luminous satellites results in a significant discrepancy between the circular velocity profiles of SPH and DM-only satellites. We find that the $v_c$ at 1 kpc predicted for satellites by DM-only simulations should be reduced by $\Delta(v_c,1kpc) \sim 0.2 v_{max, DM-only}-0.26$ km/s,  for satellites with $20 < v_{max} < 50$ km/s at infall. }

\end{itemize}

High resolution that allows simulators to limit star formation to high density 
peaks is an essential requirement to reproduce the baryonic effects in this paper.  
However, \citet{Governato2010} and \citet{Guedes2011} showed that, even at high resolution, 
if star formation is allowed to occur diffusely across the disk, no large scale 
outflows are generated.  Restricting star formation to high density 
peaks instead leads to overpressurized regions of hot gas when stars go SNe, 
leading to outflows.  These overpressurized regions expand faster than the local 
dynamical time.  When this occurs in the central $\sim$1 kpc, the potential 
flattens as this hot gas expands, leading to an irreversible expansion of the DM orbits 
\citep{Pontzen2012}.  Restricting star formation to high density peaks is 
comparable to allowing stars to only form in giant molecular clouds, rather 
than across the entire disk at any given time.  The simulations used in 
this work are the first at these high resolutions to include metal line 
cooling \citep{Shen2010} and a presciption for self-shielding of cold gas, 
allowing star formation to be tied to the shielded regions where H$_2$ can 
form \citep{Christensen2012}.  

It is important to note that 
simulations that do not resolve the effect of feedback at high densities 
will be unable to reproduce the results of this paper.  However, the feedback 
model employed in this paper has been shown to match the observed mass -- 
metallicity relation for galaxies as a function of redshift \citep{Brooks2007, 
Maiolino2008}, the baryonic Tully-Fisher relationship (Christensen et al. 
in prep.), the size -- luminosity relation of galaxy disks \citep{Brooks2011}, 
the $z=0$ stellar mass to halo mass relation \citep{Munshi2012}, and the 
central mass as a function of stellar mass for galaxies in the luminosity 
range in this paper \citep{Governato2012}.  This large number of successes 
in matching the observed scaling relations of galaxies lends credence to the 
particular feedback model employed in this paper to study satellite galaxies.

In addition to reproducing the above scaling relations, the star formation 
and feedback model used in this work has been used to simulate bulgeless dwarf 
disk galaxies \citep{Governato2010}.  Importantly, the processes that lead to 
bulgeless galaxies also transform cuspy DM density profiles into cored profiles, 
leading these simulations to match the central dark matter densities 
derived by the {\sc THINGS} and {\sc Little THINGS} surveys \citep{Oh2011, 
Governato2012}.  In other words, the simulations used in this work have been 
shown to reconcile the cusp/core problem in CDM.  In this paper, we have shown 
that this same model can alleviate the tension between the dense, massive 
satellites predicted by CDM with the observations of lower density, luminous 
dSph satellites \citep{Boylan-kolchin2012, Wolf2012, Hayashi2012}.  We note 
that the processes described in this paper also act to reduce the overall 
number of massive subhalos that exist at $z=0$, potentially solving the missing 
satellites problem in CDM.  Hence, it remains possible to resolve the small 
scale problems of CDM with a proper model for baryonic physics, and without 
invoking exotic forms of DM.  

The results presented here show that using DM-only CDM simulations to 
study the internal dynamics of luminous satellites will lead 
to erroneous results.  While it is safe to assign the most luminous 
satellites to the originally most massive halos, those massive halos 
will experience evolution that CDM DM-only runs don't account for.  
In \citet{Brooks2012}, we address how the model presented here
affects the interpretation of kinematic observations of the Milky Way's 
dSph population.

\acknowledgements
We thank the anonymous referee for comments that helped to improve this paper. We thank James Bullock, Michael Boylan-Kolchin, Jay Gallagher, Kim Venn,
Ryan Leaman, and Alan McConnachie for useful discussions.  Resources supporting
this work were provided by the NASA High-End Computing (HEC) Program through the
NASA Advanced Supercomputing (NAS) Division at Ames Research Center.  This
research was supported in part by the National Science Foundation under Grant No.
NSF PHY11-25915.  AZ acknowledges support from the Lady Davis Foundation.  AB
acknowledges support from The Grainger Foundation.  The work of AZ and AD has
been partly supported by the ISF through grant 6/08, by GIF through grant
G-1052-104.7/2009, by the DFG via German-Israeli project cooperation grant
STE1869/1-1.GE625/15-1, and by an NSF grant AST-1010033 at UCSC.  BW acknowledges
support from NSF grant AST-0908193. CC, TQ and FG were partly supported by NSF
grant AST-0908499.

\bibliographystyle{apj}
\bibliography{master} 

\appendix
\section{Numerical Convergence}\FloatBarrier
\begin{figure}[h]
\centering
\includegraphics[width=0.5\textwidth]{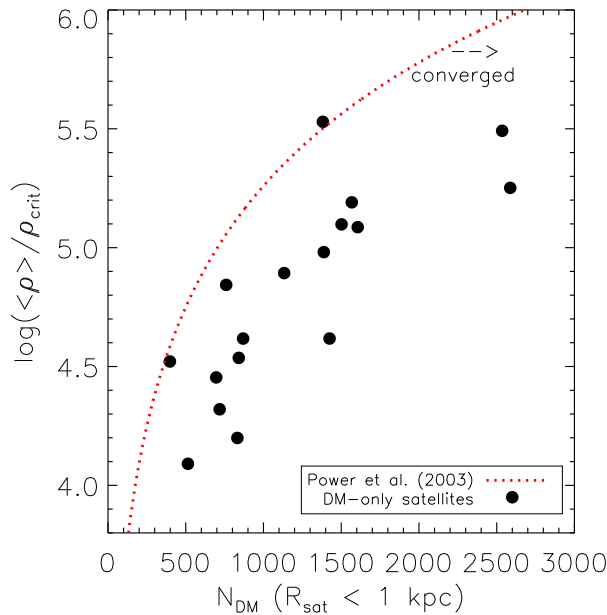}
\caption{The mean enclosed density at 1 kpc, as a function of the number of DM particles within the same radius, for DM-only satellites at infall. The density profiles of all but one of these satellites has converged by 1 kpc, according to the \citet{Power2003} criteria. The density profile of the remaining satellite has converged by 1.1 kpc.  According to this criteria, the density profiles of 15 out of the 17 satellites have already converged at $r = 0.80$ kpc.}
\label{power}
\end{figure}

\noindent The central densities of simulated halos can be artificially low due to particle relaxation when a halo contains too few particles \citep[e.g.,][]{Power2003}.  In this appendix, 
we demonstrate that the resolution of our simulations is sufficient to study the 
central density profiles of the satellites.  We verify that the creation of shallow density 
cores prior to infall is the result of physical processes (SN-driven outflows) and 
not due to low resolution in the central regions of the satellites. 

\vspace{2mm}
\noindent \citet{Power2003} have shown that the density profile of a simulated galaxy converges 
at a radius where the two-body relaxation time is larger than the Hubble time. The
radius of convergence encloses enough particles to satisfy the following criteria:
\begin{equation}
\frac{\sqrt(200)}{8} \frac{N(r)}{lnN(r)} (\frac{\bar\rho}{\rho_{crit}})^{-1/2} > 0.6
\end{equation}
where N(r) is the number of particles enclosed at a given radius, and $\bar\rho$ is 
the mean enclosed density at that radius. As we showed in this paper, SPH satellites with $M_{*} > 10^7 M_{\odot}$ at infall develop lower central densities, and since the above criterion is dependent on enclosed density,  they will meet this criterion with fewer particles than DM-only satellites with steep density profiles.  We therefore apply this 
criterion to our DM-only runs, as they will require more particles at a given radius 
to be converged.  We then require that properties of the SPH runs be measured at a 
radius where the DM-only runs converge.

\vspace{2mm}
\noindent  The Power criteria has been tested and verified for host galaxies in DM-only 
simulations, but has not yet been validated for subhalos in simulations. We therefore 
test the convergence of our DM-only runs at infall, before these satellites undergo 
mass (and hence particle) loss due to tidal stripping, but after the time that DM cores
have been created in the massive SPH runs. This allows us to test if the density profiles 
of the SPH satellites before infall are poorly resolved, which would lead to artificially 
shallow density profiles before infall. Figure~\ref{power} shows the mean enclosed density 
at 1kpc, as a function of the number of DM particles within 1 kpc, for the 17 matched 
DM-only satellites at infall. All but one of our satellites has converged at this 
radius according to the Power criteria, and we have verified that the remaining 
DM-only satellite has converged by 1.1 kpc. In fact, the density profiles of 15 out 
of the 17 satellites have already converged at $r = 0.80$ kpc, according to this criteria. 

\vspace{2mm}
\noindent At z=0, the central density profiles of the DM-only satellites have slopes 
of $-2.2 < \alpha(1$kpc) $ <-1.8$ across their mass range, consistent with 
the results of previous studies of DM-only halos \citep[e.g.,][]{Reed2005}.  
\citet{Springel2008} examined the convergence of density profiles in very high 
resolution DM-only simulations (the Aquarius simulations), and found the 
density converged at $4-6\times$ the force softening length, $\epsilon$.  
We verified, using a lower resolution DM-only run with $1.5\times$ larger 
force softening, that the density slopes in these simulations converge by 
4$\epsilon$, or 700pc.  We also confirmed that $v_c$ is 90$\%$ converged 
by 1 kpc.  The $v_c$ convergence radius is larger than for density, because 
$v_c$ is a cumulative quantity. 

\begin{figure}
\includegraphics[width=0.5\textwidth]{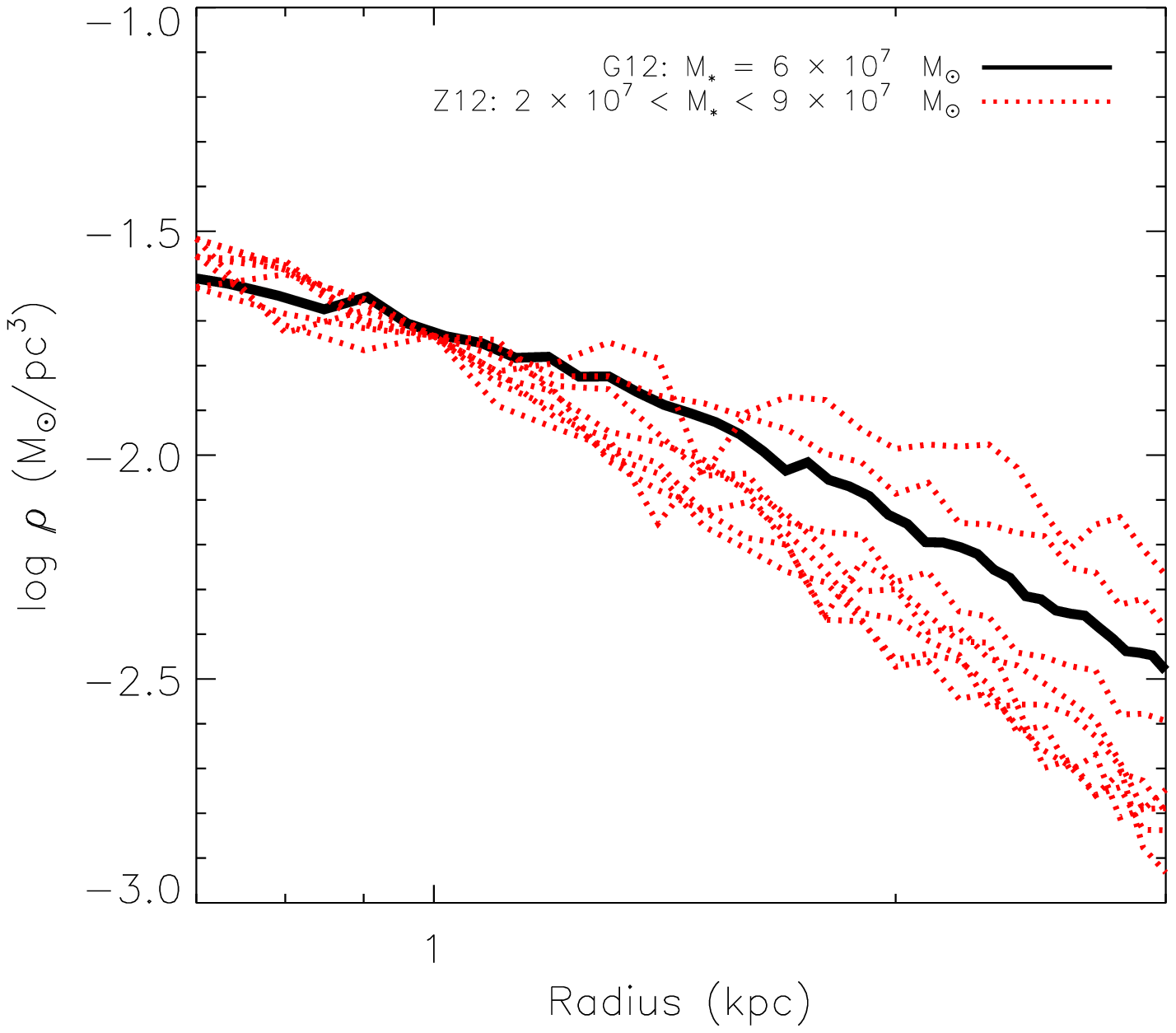}
\includegraphics[width=0.5\textwidth]{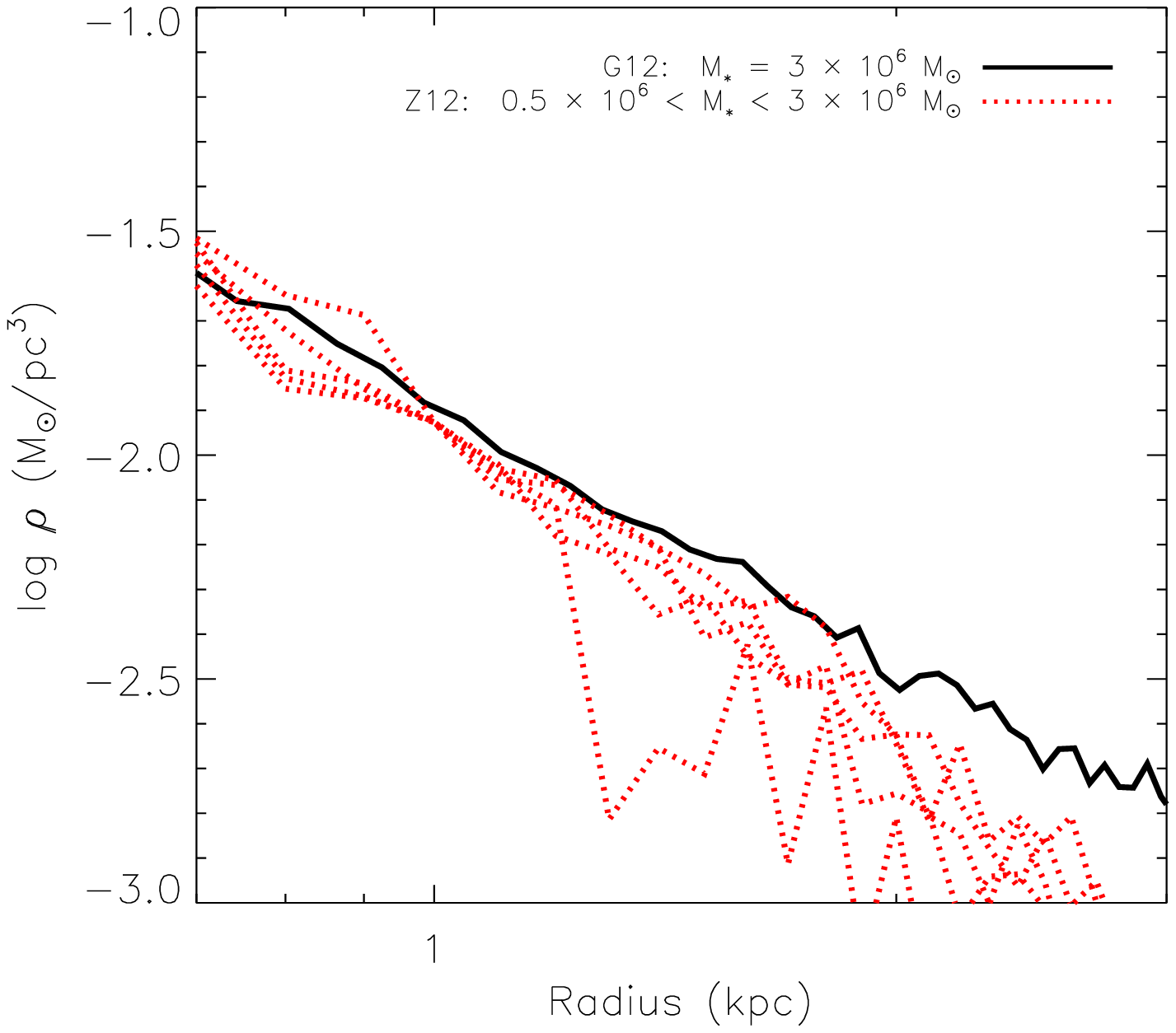}
\caption{In this figure, we show the $z=0$ DM density profiles of satellites from this paper (in red dotted lines), as well as isolated field dwarfs from Governato et al. (2012, in black solid lines).   Left Panel: The most luminous satellites in our sample; Right Panel: The least luminous galaxies in our sample. The legend lists the total stellar mass of these dwarf galaxies at $z=0$. }
\label{den.comp}
\end{figure}

\vspace{2mm}
\noindent Additionally, we can test whether the density profiles of our SPH satellites are 
in agreement with higher resolution simulation studies of dwarf galaxies. As discussed 
throughout this paper, \citet[][hereafter, G12]{Governato2012} showed that SN-driven 
outflows create DM cores in isolated field dwarfs with $M_{*} >10^7 M_{\odot}$. These 
isolated dwarf simulations were run with the same star formation and feedback scheme 
as the simulations used in this paper, but at $8 \times$ higher mass resolution and 
twice the force resolution. Two field dwarfs in the G12 sample have comparable stellar 
masses to the satellites studied here, one with $M_{*}  =  6 \times 10^7 M_{\odot}$ 
and one with $M_{*} = 3 \times 10^6 M_{\odot}$. 
Figure~\ref{den.comp} shows the DM density profiles of our most luminous satellites 
(which initially contain the highest number of particles and are therefore the best 
resolved, left panel) and our least luminous satellites (containing the fewest 
particles and should be least resolved, right panel) at z=0. 
We have normalized the density of the galaxies in each panel at 1 kpc, such that 
all galaxies have the same $\rho(1$ kpc), for a clear comparison of their DM 
density slopes.  Two results are evident in Figure~\ref{den.comp}.   
First, the low-luminosity dwarfs in the right panel all have steep DM density 
profiles, while the more luminous dwarfs in the left panel have shallower DM 
profiles within 1 kpc. Second, the satellite density slopes in each mass range 
interior to 1 kpc are comparable to the higher resolution G12 results (profiles 
at radii larger than 1 kpc can be steeper in the satellites due to the tidal 
stripping they have experienced).  We verified that the central density profiles of these 
two samples at infall also matches the $z=0$ G12 slopes, indicating little 
evolution in the central density profiles between infall and $z=0$.  The comparable slopes 
of the satellite profiles to the more highly resolved field dwarfs is further 
evidence that the density profiles of our satellites have converged by $0.7$ kpc.

\end{document}